\documentclass[11pt,english]{article}
\usepackage[T1]{fontenc}
\usepackage[latin9]{inputenc}
\usepackage{mathrsfs}
\usepackage{amsmath}
\usepackage{amssymb}
\usepackage{esint}

\usepackage{a4wide}

\makeatother

\usepackage{babel}

\begin{document}

\title{Quantum Field Theory of Classically Unstable Hamiltonian Dynamics}

\author{Y. Strauss${}^{1,2,a)}$, L.P. Horwitz${}^{3,4,1}$, J. Levitan${}^1$,
A. Yahalom${}^5$}
\maketitle
\begin{description}
\item [{$^{1}$}] Department of Physics, Ariel University, Ariel 40700,
Israel
\item [{$^{2}$}] Department of Mathematics, Ben-Gurion University of the
Negev, Be'er Sheva 84105, Israel
\item [{$^{3}$}] School of Physics and Astronomy, Raymond and Beverly
Sackler Faculty of Exact Sciences, Tel-Aviv university, Tel-Aviv 69978,
Israel 
\item [{$^{4}$}] Department of Physics, Bar-Ilan University, Ramat-Gan
52900, Israel
\item [{$^{5}$}] Department of Electrical and Electronic Engineering,
Ariel 40700, Israel\end{description}
%
\setcounter{footnote}{1}
\footnotetext{e-mail: yossef.strauss@gmail.com}
\begin{abstract}
We study a class of dynamical systems for which the motions can be
described in terms of geodesics on a manifold (ordinary potential
models can be cast into this form by means of a conformal map). It
is rigorously proven that the geodesic deviation equation of Jacobi,
constructed with a second covariant derivative, is unitarily equivalent
to that of a parametric harmonic oscillator, and we study the second
quatization of this oscillator. The excitations of the Fock space
modes correspond to the emission and absorption of quanta into the
dynamical medium, thus associating unstable behavior of the dynamical
system with calculable fluctuations in an ensamble with possible thermodynamic
consequences. 
\end{abstract}

\section{Introduction\label{sec:introduction}}

There are dynamical systems which can be described by an evolution
generated by a geometrical Hamiltonian of the form

\begin{equation}
H(x,p)=\frac{1}{2m}g^{ij}(x)p_{i}p_{j}\,.\label{eq:geometric_hamiltonian-1}
\end{equation}
An example of such a system is the geodesic motion of general relativity
which can be understood as the application of Hamilton's equations
to a geometrical Hamiltonian in the framework of a symplectic embedding
of the spacetime manifold \cite{key-16}. Another example is the description
of the motion of an electron near the boundaries of a Brillouin zone
where the inverse mass matrix of the electron plays the role of a
metric \cite{key-18}. Geometric Hamiltonians of the form of Eq. (\ref{eq:geometric_hamiltonian-1})
also naturally appear in procedures of geometrization of Newtonian
dynamics. Starting with Hamiltonian of the form 

\begin{equation}
H=\frac{p^{2}}{2m}+V(x)\label{eq:standard_hamitonian}
\end{equation}
two common geometrization schemes involve the introduction of the
Jacobi metric \cite{key-15} or the Eisenhart metric \cite{key-12}.
Moreover, it has recently been shown \cite{key-1} that the dynamics
generated by a Hamiltonian of the form of Eq. (\ref{eq:standard_hamitonian})
can be represented, by means of a conformal map, in terms of the dynamics
of a geodesic flow on a manifold generated by a geometric Hamiltonian
of the form of Eq. (\ref{eq:geometric_hamiltonian-1}). The flow generated
according to Hamilton's equations by the geometric Hamiltonian Eq.
(\ref{eq:geometric_hamiltonian-1}) is described by the geodesic equation
\begin{equation}
\ddot{x}^{i}+\Gamma_{jk}^{i}\dot{x}^{j}\dot{x}^{k}=0\label{eq:geodesic_equation}
\end{equation}
where $\dot{x}^{i}$ is $dx^{i}/ds$ (in the geometrical picture $s$
is understood as the arc length parameter).

The stability of the geodesic flow is locally determined by the geodesic
deviation equation
\begin{equation}
\frac{D^{2}\xi^{i}}{dt^{2}}+R_{jkl}^{i}\dot{x}^{j}\dot{x}^{k}\xi^{l}=0\label{eq:geodesic_dev_eqn_component}
\end{equation}
 where $D/dt$ is covariant derivative and $\xi^{i}$ are the components
of the geodesic deviation vector $\xi^{i}(t)=\frac{\partial x^{i}(\alpha,t)}{\partial\alpha}|_{\alpha=0}$
where $\alpha$ is the parameter for a family of geodesics in the
neighborhood of the coordinates $x^{i}(t)$ of a point on a geodesic
defined by Eq. (\ref{eq:geodesic_equation}). It was demonstrated
in a large number of cases \cite{key-9,key-10}, that the stability
of this geodesic motion with conformal metric is related to stability
of the motion generated by the Hamiltonian in Eq. (\ref{eq:standard_hamitonian})
(as seen through Lyapunov exponents and Poincare plots). 

Viewing the geodesic deviation equation, Eq. (\ref{eq:geodesic_dev_eqn_component}),
as an oscillator equation \cite{key-2,key-3}, one can understand
the local stability of the geodesic motion on the manifold in terms
of the stability of the associated oscillator. Casseti, Clementi and
Pettini \cite{key-4} discuss the idea that this oscillator is essentially
parametric due to curvature fluctuations on manifolds whose natural
motions are geodesic motions and associate dynamical instability of
Hamiltonian systems with parametric instability of the oscillator.

In Sec. \ref{sec:parametric_osc_dyn_sys} we discuss an exact representation
of the geodesic deviation in terms of a parametric oscillator equation.
Under an adiabatic assumption, the motion generated by this equation
can be embedded into a unitary evolution in a Hilbert space through
a process of dilation (see Appendix A) which provides additional degrees
of freedom corresponding to a dynamical environment. The oscillatory,
stable (decaying) and unstable behavior of the oscillator correspond
in this embedding to the effective interaction of the system with
a dynamical environment.

The dynamics of the states in the second quantization of the embedding
Hilbert space represents the interaction of the system with the dynamical
environment reflecting the stability properties of the motion generated
by the Hamiltonian in Eq. (\ref{eq:geometric_hamiltonian-1}). This
phenomenon is a remarkable property of complex systems which lends
itself to a rigorous description in terms of the procedure of dilation
and second quantization.

The rest of the paper is organized as follows: In Section \ref{sec:parametric_osc_dyn_sys}
we prepare the ground for the analysis of the stability of the dynamics
generated by the geodesic deviation equation by showing that it can
be mapped unitarily into a parametric oscillator equation defined
on a fixed (finite dimensional) Hilbert space. This is done in Subsection
\ref{sub:parametric_oscillator_rep}. By the unitarity of the mapping
involved, the dynamics of the parametric oscillator representation
is completely equivalent to that of the original geodesic deviation
equation. However, the parametric oscillator representation does not
separate stable from unstable behavior, as is also true for the original
geodesic deviation equation (since it is a second order equation).
For this we introduce in Subsection \ref{sub:dynamical_system_rep},
via a second unitary mapping, a dynamical system representation which,
by composition of mappings, is again exactly equivalent to the geodesic
deviation equation. In the dynamical system representation we make
an adiabatic approximation enabling us to identify stable and unstable
subspaces for the evolution of the geodesic deviation. In Subsection
\ref{sub:isometric_dilation_stable_unstable_evol}, using the same
adiabatic approximation, we identify contractive semigroups corrsesponding
to the restriction of the evolution of the geodesic deviation onto
the stable and unstable subspaces and apply a dilation procedure which
effectively extends the system and provides degrees of freedom (i.e.,
appropriate subspaces of the dilation Hilbert space) which later on,
following an application of a procedure of second quantization in
Subsection \ref{sub:second_quantization}, are interpreted as a dynamical
environment inducing stability or instality of the evolution of the
geodesic deviation. It is to be noted that a natural setting for the
application of the dilation (and second quantization) procedure are
function spaces defined along a geodesic (i.e., the geodesic with
respect to which the geodesic deviation is defined). These function
spaces, and the mapping of the dilation structure onto them, are described
in Subsection \ref{sub:isometric_dilations_on_geodesics}. Conclusions
and some remarks on possible avenues of further progress along the
lines introduced in the present paper are given in Section \ref{sec:conclusions}.
Appendix A provides a short description of a procedure for the construction
of unitary and isometric dilations of continuous, one parameter, contractive
semigroups. A thorough treatment of unitary and isometric dilations
of such semigroups is found in \cite{key-5}.

\section{Parametric oscillator and dynamical system representations of the
geodesic deviation equation\label{sec:parametric_osc_dyn_sys}}

\subsection{Parametric oscillator representation of the geodesic deviation equation\label{sub:parametric_oscillator_rep}}

In the following we introduce the basic mathematical definitions and
tools for the description of the dynamical properties of the geodesic
deviation associated with a geodesic flow on a Riemannian manifold.
These tools apply directly to the geodesic flow generated by the geometric
Hamiltonian in Eq. (\ref{eq:geometric_hamiltonian-1}), described
in the previous section.

Consider an $n$ dimensional Riemannian manifold $\mathcal{M}$ and
a geodesic curve $\gamma$ in $\mathcal{M}$. Let $T\mathcal{M}$
be the tangent bundle of $\mathcal{M}$ and let $T_{p}\mathcal{M}$
be the tangent space to $\mathcal{M}$ at the point $p\in\mathcal{M}$.
Denote further the part of $T\mathcal{M}$ over the geodesic $\gamma$
by $T_{\gamma}\mathcal{M}$. Let $p_{0}\in\gamma$ be an arbitrary
point on the geodesic $\gamma$ and consider an arc length parametrization
$\gamma(\cdot)\,:\,\mathbb{R}\mapsto\mathcal{M}$ of $\gamma$ 
\begin{equation}
\gamma(s)=\exp_{p_{0}}(s\mathbf{v})\label{eq:geodesic_arc_length_para}
\end{equation}
where $\mathbf{v}\in T_{p_{0}}\mathcal{M}$ is a unit tangent vector
to $\gamma$ at the point $p_{0}$. Thus, for $s\geq0$, $\gamma(s)$
is a point at arc length distance $s$ from the point $p_{0}$ along
the geodesic that starts at $p_{0}$ and has tangent vector $\mathbf{v}$
at $p_{0}$, and for $s<0$, $\gamma(s)$ is a point at arc length
$\left|s\right|$ from the point $p_{0}$ along the geodesic that
starts at $p_{0}$ and has tangent vector $(-\mathbf{v})$ at $p_{0}$.
We assume that $\mathcal{M}$ is geodesically complete, i.e., all
of the geodesics starting at an arbitrary point in $\mathcal{M}$
can be continued indefinitely.

Let $g$ be the metric tensor of $\mathcal{M}$ and denote the scalar
product in the tangent space $T_{p}\mathcal{M}$ by $\langle\cdot,\,\cdot\rangle_{T_{p}\mathcal{M}}$,
i.e., if $\mathbf{X},\mathbf{Y}\in T_{p}\mathcal{M}$ are two tangent
vectors at the point $p$ then we have
\[
\langle\mathbf{X},\mathbf{Y}\rangle_{T_{p}\mathcal{M}}=g(\mathbf{X},\mathbf{Y})
\]
We assume that the connection on $\mathcal{M}$ is given by the Christoffel
symbols. In this case we have a compatibility of the covariant derivative
over $\mathcal{M}$ with its metric structure. If we denote by $\frac{\nabla}{dt}$
the covariant derivative along a smooth curve $\gamma$ in $\mathcal{M}$,
parametrized by $t$, and if $\mathbf{X}(t)$ and $\mathbf{Y}(t)$
are two smooth vector valued functions defined along $\gamma,$ we
then have

\begin{equation}
\frac{d}{dt}\langle\mathbf{X}(t),\mathbf{Y}(t)\rangle_{T_{\gamma(t)}\mathcal{M}}=\left\langle \frac{\nabla\mathbf{X}(t)}{dt},\ \mathbf{Y}(t)\right\rangle _{T_{\gamma(t)}\mathcal{M}}+\left\langle \mathbf{X}(t),\,\frac{\nabla\mathbf{Y}(t)}{dt}\right\rangle _{T_{\gamma(t)}\mathcal{M}}\label{eq:connection_metric_compatibility}
\end{equation}
In particular, if we obtain the two functions $\mathbf{X}(t)$ and
$\mathbf{Y}(t)$ by parallel transport along $\gamma$ of two vectors
$\mathbf{X},\mathbf{Y}\in T_{p_{0}}\mathcal{M}$, given in the tangent
space of a single point $p_{0}\in\gamma(0)$, then by definition of
parallel transport we have $\frac{\nabla\mathbf{X}(t)}{dt}=0$ and
$\frac{\nabla\mathbf{Y}(t)}{dt}=0$ and so in this case, using Eq.
(\ref{eq:connection_metric_compatibility}), we have that $\frac{d}{dt}\langle\mathbf{X}(t),\mathbf{Y}(t)\rangle_{T_{\gamma(t)}\mathcal{M}}=0$.
Let us denote by $\phi(s)\mathbf{X}$ the parallel transport of a
vector $\mathbf{X}\in T_{\gamma(t)}\mathcal{M}$ to the tangent space
$T_{\gamma(t+s)}\mathcal{M}$ along the curve $\gamma$. Then $\phi(\cdot)$
defines a continuous surjective mapping $\phi(\cdot)\,:\, T_{\gamma}\mathcal{M}\mapsto T_{\gamma}\mathcal{M}$
and, under the compatibility assumption above, we have
\begin{equation}
\langle\phi(s)\mathbf{X},\phi(s)\mathbf{Y}\rangle_{T_{\gamma(t+s)}\mathcal{M}}=\langle\mathbf{X},\mathbf{Y}\rangle_{T_{\gamma(t)}\mathcal{M}},\qquad\mathbf{X},\mathbf{Y}\in T_{\gamma(t)}\mathcal{M}\label{eq:parallel_trans_unitary_map}
\end{equation}
i.e., parallel transport is a unitary mapping between the tangent
spaces along $\gamma$. 

Let $\gamma$ be a geodesic curve parametrized by arc length parameter
$s$ as in Eq. (\ref{eq:geodesic_arc_length_para}). Let $C^{1}(\mathbb{R};\, T_{p_{0}}\mathcal{M})$
be the space of all $C^{1}$ vector valued functions defined on the
real axis $\mathbb{R}$ and taking values in the vector space $T_{p_{0}}\mathcal{M}$.
Let $C^{1}(\mathbb{R};\, T_{\gamma}\mathcal{M})$ be the space of
all $C^{1}$ vector valued functions defined on the real axis $\mathbb{R}$,
taking values in $T_{\gamma}\mathcal{M}$ and satisfying the condition
that 
\[
\mathbf{X}(\cdot)\in C^{1}(\mathbb{R};\, T_{\gamma}\mathcal{M})\ \Rightarrow\ \mathbf{X}(s)\in T_{\gamma(s)}\mathcal{M},\ \forall s\in\mathbb{R}.
\]
Let $\mathbf{X}(\cdot),\mathbf{Y}(\cdot)\in C^{1}(\mathbb{R};\, T_{\gamma}\mathcal{M})$
be arbitrary vector valued functions. Denoting $\tilde{\mathbf{X}}(s)=\phi^{-1}(s)\mathbf{X}(s)$
and $\tilde{\mathbf{Y}}(s)=\phi^{-1}(s)\mathbf{Y}(s)$ we have $\tilde{\mathbf{X}}(s),\,\tilde{\mathbf{Y}}(s)\in T_{p_{0}}\mathcal{M}$
and so the functions $\tilde{\mathbf{X}}(\cdot)$, $\tilde{\mathbf{Y}}(\cdot)$
belong to $C^{1}(\mathbb{R};\, T_{p_{0}}\mathcal{M})$. By the unitarity
of the parallel transport, following from Eq. (\ref{eq:parallel_trans_unitary_map}),
for such arbitrary vector valued functions $\mathbf{X}(\cdot),\mathbf{Y}(\cdot)\in C^{1}(\mathbb{R};\, T_{\gamma}\mathcal{M})$
we have 
\begin{equation}
\left\langle \mathbf{X}(s),\,\mathbf{Y}(s)\right\rangle _{T_{\gamma(s)}\mathcal{M}}=\left\langle \phi^{-1}(s)\mathbf{X}(s),\phi^{-1}(s)\mathbf{Y}(s)\right\rangle _{T_{p_{0}}\mathcal{M}}=\left\langle \mathbf{\tilde{X}}(s),\mathbf{\tilde{Y}}(s)\right\rangle _{T_{p_{0}}\mathcal{M}}\label{eq:parallel_trans_unitary_map_2}
\end{equation}
Thus we may define a mapping $W_{\gamma}\,:\, C^{1}(\mathbb{R};\, T_{p_{0}}\mathcal{M})\mapsto C^{1}(\mathbb{R};\, T_{\gamma}\mathcal{M})$
by
\begin{equation}
[W_{\gamma}\tilde{\mathbf{X}}(\cdot)](s):=\phi(s)\tilde{\mathbf{X}}(s),\quad\tilde{\mathbf{X}}(\cdot)\in C^{1}(\mathbb{R};\, T_{p_{0}}\mathcal{M})\label{eq:w_gamma_def}
\end{equation}
and according to Eq. (\ref{eq:parallel_trans_unitary_map_2}) we have
\begin{equation}
\left\langle \mathbf{\tilde{X}}(s),\mathbf{\tilde{Y}}(s)\right\rangle _{T_{p_{0}}\mathcal{M}}=\left\langle [W_{\gamma}\tilde{\mathbf{X}}(\cdot)](s),\,[W_{\gamma}\tilde{\mathbf{Y}}(\cdot)](s)\right\rangle _{T_{\gamma(s)}\mathcal{M}}\label{eq:w_gamma_local_unitarity}
\end{equation}
Note also that the inverse of $W_{\gamma}$ is given by
\[
[W_{\gamma}^{-1}\mathbf{X}(\cdot)](s):=\phi^{-1}(s)\mathbf{X}(s),\ \mathbf{X}(\cdot)\in C^{1}(\mathbb{R};\, T_{\gamma}\mathcal{M}).
\]

Now, let $\mathbf{Y}(\cdot)\in C^{1}(\mathbb{R};\, T_{\gamma}\mathcal{M})$
be arbitrary and let $\mathbf{X}(\cdot)\in C^{1}(\mathbb{R};\, T_{\gamma}\mathcal{M})$
be a vector valued function defined by $\mathbf{X}(s)=\phi(s)\mathbf{X}_{0}$,
where $\mathbf{X}_{0}\in T_{\gamma(0)}\mathcal{M}=T_{p_{0}}\mathcal{M}$
,i.e., $\mathbf{X}(\cdot)$ is obtained by the parallel transport
along $\gamma$ of a fixed vector $\mathbf{X}_{0}\in T_{p_{0}}\mathcal{M}$.
Using Eq. (\ref{eq:connection_metric_compatibility}) we then have
\[
\frac{d}{ds}\langle\mathbf{X}(s),\mathbf{Y}(s)\rangle_{T_{\gamma(s)}\mathcal{M}}=\left\langle \mathbf{X}(s),\,\frac{\nabla\mathbf{Y}(s)}{ds}\right\rangle _{T_{\gamma(s)}\mathcal{M}}=\left\langle \phi(s)\mathbf{X}_{0},\,\frac{\nabla\mathbf{Y}(s)}{ds}\right\rangle _{T_{\gamma(s)}\mathcal{M}}
\]
and hence
\begin{multline*}
\left\langle \phi(s)\mathbf{X}_{0},\,\frac{\nabla\mathbf{Y}(s)}{ds}\right\rangle _{T_{\gamma(s)}\mathcal{M}}=\frac{d}{ds}\langle\mathbf{X}(s),\mathbf{Y}(s)\rangle_{T_{\gamma(s)}\mathcal{M}}=\frac{d}{ds}\left\langle \phi^{-1}(s)\mathbf{X}(s),\phi^{-1}(s)\mathbf{Y}(s)\right\rangle _{T_{p_{0}}\mathcal{M}}=\\
=\frac{d}{ds}\left\langle \mathbf{X}_{0},\phi^{-1}(s)\mathbf{Y}(s)\right\rangle _{T_{p_{0}}\mathcal{M}}=\left\langle \mathbf{X}_{0},\frac{d}{ds}[\phi^{-1}(s)\mathbf{Y}(s)]\right\rangle _{T_{p_{0}}\mathcal{M}}
\end{multline*}
Furthermore, since
\[
\left\langle \phi(s)\mathbf{X}_{0},\,\frac{\nabla\mathbf{Y}(s)}{ds}\right\rangle _{T_{\gamma(s)}\mathcal{M}}=\left\langle \mathbf{X}_{0},\,\phi^{-1}(s)\frac{\nabla\mathbf{Y}(s)}{ds}\right\rangle _{T_{p_{0}}\mathcal{M}}
\]
we get
\[
\left\langle \mathbf{X}_{0},\,\phi^{-1}(s)\frac{\nabla\mathbf{Y}(s)}{ds}\right\rangle _{T_{p_{0}}\mathcal{M}}=\left\langle \mathbf{X}_{0},\frac{d}{ds}[\phi^{-1}(s)\mathbf{Y}(s)]\right\rangle _{T_{p_{0}}\mathcal{M}}
\]
and since $\mathbf{X}_{0}\in T_{p_{0}}\mathcal{M}$ is arbitrary,
we find that
\begin{equation}
\phi^{-1}(s)\frac{\nabla\mathbf{Y}(s)}{ds}=\frac{d}{ds}[\phi^{-1}(s)\mathbf{Y}(s)]\label{eq:der_covariant_der_intertwine}
\end{equation}
If we set $\tilde{\mathbf{Y}}(s)=\phi^{-1}(s)\mathbf{Y}(s)$ and write
in short form $\mathbf{Y}\equiv\mathbf{Y}(\cdot)$, $\tilde{\mathbf{Y}}\equiv\tilde{\mathbf{Y}}(\cdot)$
we can write Eq. (\ref{eq:der_covariant_der_intertwine}) in the form
\begin{equation}
\frac{\nabla}{ds}(W_{\gamma}\tilde{\mathbf{Y}})(s)=\left(W_{\gamma}\frac{d\tilde{\mathbf{Y}}}{ds}\right)(s)\label{eq:der_covariant_der_intertwine_2}
\end{equation}
The mapping $W_{\gamma}$ then intertwines covariant derivative with
ordinary derivative.

Next we extend the mapping $W_{\gamma}$ to more general tensor valued
functions. Denote by $\Lambda_{p}(l,\, k)$ the set of all $\left(l,k\right)$-type
tensors defined at a point $p\in\mathcal{M}$, i.e., the set of all
$k$ times covariant and $l$ times contravriant tensors at $p$.
Let $\Lambda_{\gamma}(l,\, k)$ be the bundle of all $(l,k)$-type
tensors defined at all points of a smooth curve $\gamma$, i.e., $\Lambda_{\gamma}(l,\, k):=\cup_{p\in\gamma}\Lambda_{p}(l,k)$.
Note that both notations, $\Lambda_{p}(l,\, k)$ and $\Lambda_{\gamma}(l,\, k)$,
ignore the ordering of covariant and contravariant arguments of the
tensors with which we are concerned. This ordering is determined by
the context of our work. Let $C^{1}(\mathbb{R};\,\Lambda_{\gamma}(l,\, k))$
be the space of all $C^{1}$ tensor valued functions, of type $(l,k)$,
defined on a smooth curve $\gamma$. If $T(\cdot)$ is such a tensor
valued function we denote by $T_{\gamma(t)}$ its value at the point
$\gamma(t)\in\gamma$, so that $T_{\gamma(t)}\in\Lambda_{\gamma(t)}(l,k)$.
We shall usually use the abreviated notation $T\equiv T(\cdot)$.
We extend our notation of parallel transport and denote by $\phi(t)T$
the parallel transport along $\gamma$ of a tensor $T_{\gamma(s)}\in\Lambda_{\gamma(s)}(l,k)$
to the space $\Lambda_{\gamma(t+s)}(l,k)$. Now let $\gamma$ be a
geodesic parametrized by arc length as in Eq. (\ref{eq:geodesic_arc_length_para})
and let $p_{0}=\gamma(0)$. Let $C^{1}(\mathbb{R};\,\Lambda_{p_{0}}(l,k))$
be the space of all $(l,k)$-type tensor valued functions defined
on the real axis $\mathbb{R}$ and taking values in $\Lambda_{p_{0}}(l,k)$.
We define a mapping $W_{\gamma}\,:\, C^{1}(\mathbb{R};\,\Lambda_{p_{0}}(l,k))\mapsto C^{1}(\mathbb{R};\,\Lambda_{\gamma}(l,\, k))$
such that, for each tensor valued function $\tilde{T}(\cdot)\in C^{1}(\mathbb{R};\,\Lambda_{p_{0}}(l,k))$
we have 

\begin{equation}
[W_{\gamma}\tilde{T}](s)\,:=\phi(s)\tilde{T}(s)\label{eq:w_gamma_extended_def}
\end{equation}
If $T\in C^{1}(\mathbb{R};\,\Lambda_{\gamma}(l,\, k))$ is a tensor
valued function defined on the geodesic $\gamma$ the action of the
inverse of $W_{\gamma}$ on $T$ is given by
\begin{equation}
[W_{\gamma}^{-1}T](s)=\phi(-s)T_{\gamma(s)}\label{eq:w_gamma_extended_inverse}
\end{equation}

Now we turn to consider the geodesic deviation equation. Let $\gamma$
be a geodesic on $\mathcal{M}$. We vary $\gamma$ into a family $\left\{ \gamma_{\alpha}\right\} _{\alpha\in I}$
of geodesics depending on a parameter $\alpha\in(-\delta,\delta)=I$,
with $\gamma_{0}=\gamma$. We consider all of the geodesics in the
family to be parametrized by the arc length parameter $s$, as in
Eq. (\ref{eq:geodesic_arc_length_para}). Thus, in terms of coordinates
$x=(x^{1},\ldots,x^{n})$ in a coordinate patch in $\mathcal{M}$,
the coordinates along $\gamma$ are $x(s,0)$ and the coordinates
along $\gamma_{\alpha}$ are $x(s,\alpha)$. If we denote symbolically
the points on the surface parametrized by $\alpha$ and $s$ by $\vec{\mathbf{x}}(\alpha,s)$
then the geodesic deviation vector is defined to be 
\[
\mathbf{J}(s):=\left.\frac{\partial\vec{\mathbf{x}}(\alpha,s)}{\partial\alpha}\right|_{\alpha=0}
\]
and if the coordinate basis vectors associated with the coordinates
$x$ are denoted by $\vec{e}_{i}\equiv\partial_{i}$, $i=1,\ldots,n$,
then we have
\[
\mathbf{J}(s)=\left.\frac{\partial x^{i}(s,\alpha)}{\partial\alpha}\right|_{\alpha=0}\vec{e_{i}}(s)=\xi^{i}(s)\vec{e}_{i}(s)
\]
where $\xi^{i}(s)=\left.\frac{\partial x^{i}(s,\alpha)}{\partial\alpha}\right|_{\alpha=0}$.
Since
\[
x^{i}(s,\alpha)-x^{i}(s,0)=\alpha\left(\left.\frac{\partial x^{i}(s,\alpha)}{\partial\alpha}\right|_{\alpha=0}\right)+O(\alpha^{2})
\]
then $\alpha\mathbf{J}(s)$ is a vector representing the linear approximation
to the separation between the geodesic $\gamma$ and the geodesic
$\gamma_{\alpha}$. Let $\mathbf{X},\mathbf{Y},\mathbf{Z}\in T_{p}\mathcal{M}$
be vectors and let $R_{p}(\mathbf{X},\mathbf{Y})\,:\, T_{p}\mathcal{M}\mapsto T_{p}\mathcal{M}$
be the curvature transformation at the point $p$, i.e., $R_{p}(\mathbf{X},\mathbf{Y})$
is a linear transformation with matrix elements $[R_{p}(\mathbf{X},\mathbf{Y})]_{\, j}^{i}=R_{jkl}^{i}X^{k}Y^{l}$
so that 
\[
R_{p}(\mathbf{X},\mathbf{Y})\mathbf{Z}=(R_{jkl}^{i}X^{k}Y^{l}Z^{j})\vec{e}_{i}=(R_{jkl}^{i}X^{k}Y^{l}Z^{j})\partial_{i}
\]
The quantities $R_{jkl}^{i}$ are the components of the Riemann curvature
tensor (evaluated at $p$). Note also that for $\mathbf{W}\in T_{p}\mathcal{M}$
we have $\left\langle R_{p}(\mathbf{X},\mathbf{Y})\mathbf{Z},\,\mathbf{\mathbf{W}}\right\rangle _{T_{p}\mathcal{M}}=R_{jkl}^{i}X^{k}Y^{l}Z^{j}W_{i}$,
where $W_{i}=g_{ij}W^{j}$. Using the above notation for the curvature
transformation, the geodesic deviation equation has the form \cite{key-13}
\begin{equation}
\frac{\nabla^{2}\mathbf{\mathbf{J}}(s)}{ds^{2}}+R_{\gamma(s)}(\mathbf{J}(s),\mathbf{T}(s))(\mathbf{T}(s))=\mathbf{0}\label{eq:geodesic_dev_eqn}
\end{equation}
where $R_{\gamma(s)}$ is the curvature tensor at the point $\gamma(s)\in\gamma$,
$\mathbf{J}(s)$ is the geodesic deviation vector and $\mathbf{T}(s)\equiv\mathbf{T}_{\gamma(s)}$
is the tangent vector to $\gamma$ at the point $\gamma(s)$. The
component representation of this equation is, of course, Eq. (\ref{eq:geodesic_dev_eqn_component})
above. Take a vector $\mathbf{X}_{0}\in T_{p_{0}}\mathcal{M}$ and
parallel transport it along the geodesic $\gamma$ to obtain a vector
valued function $\mathbf{X}(\cdot)$ given by $\mathbf{X}(s)=\phi(s)\mathbf{X}_{0}$.
We have, of course,
\[
\left\langle \mathbf{X}(s),\,\frac{\nabla^{2}\mathbf{J}(s)}{ds^{2}}+R_{\gamma(s)}(\mathbf{J}(s),\mathbf{T}(s))\mathbf{T}(s)\right\rangle _{T_{\gamma(s)}\mathcal{M}}=0
\]
By the definition of parallel transport of tensors along $\gamma$
we obtain
\begin{multline}
0=\left\langle \mathbf{X}(s),\,\frac{\nabla^{2}\mathbf{J}(s)}{ds^{2}}+R_{\gamma(s)}(\mathbf{J}(s),\mathbf{T}(s))\mathbf{T}(s)\right\rangle _{T_{\gamma(s)}\mathcal{M}}=\\
=\left\langle W_{\gamma}^{-1}\mathbf{X}(s),\, W_{\gamma}^{-1}\left(\frac{\nabla^{2}\mathbf{J}(s)}{ds^{2}}+R_{\gamma(s)}(\mathbf{J}(s),\mathbf{T}(s))\mathbf{T}(s)\right)\right\rangle _{T_{p_{0}}\mathcal{M}}=\\
=\left\langle \mathbf{X}_{0},\,\frac{d^{2}(W_{\gamma}^{-1}\mathbf{J})(s)}{ds^{2}}+(W_{\gamma}^{-1}R_{\gamma(s)})(W_{\gamma}^{-1}\mathbf{J})(s),(W_{\gamma}^{-1}\mathbf{T})(s))(W_{\gamma}^{-1}\mathbf{T})(s)\right\rangle _{T_{p_{0}}\mathcal{M}}\label{eq:geodesic_dev_to_parametric_osc_1}
\end{multline}
Recall that the geometrical form of the geodesic equation for $\gamma$
is
\begin{equation}
\frac{\nabla\mathbf{T}(s)}{ds}=\mathbf{0}\label{eq:geodesic_eqn_geometric_form}
\end{equation}
and Eq. (\ref{eq:geodesic_equation}) is the component representation
of Eq. (\ref{eq:geodesic_eqn_geometric_form}). This implies that
$(W_{\gamma}^{-1}\mathbf{T})(s)=\phi(-s)\mathbf{T}_{\gamma(s)}=\phi(-s)\mathbf{T}(s)=\mathbf{T}_{0}$,
where $\mathbf{T}_{0}\in T_{p_{0}}\mathcal{M}$ is the tangent vector
to $\gamma$ at the point $p_{0}=\gamma(0)$. Hence Eq. (\ref{eq:geodesic_dev_to_parametric_osc_1})
can be written in the form
\[
0=\left\langle \mathbf{X}_{0},\,\frac{d^{2}(W_{\gamma}^{-1}\mathbf{J})(s)}{ds^{2}}+(W_{\gamma}^{-1}R_{\gamma(s)})(W_{\gamma}^{-1}\mathbf{J})(s),\mathbf{T}_{0})(\mathbf{T}_{0})\right\rangle _{T_{p_{0}}\mathcal{M}}
\]
Denoting $\tilde{\mathbf{J}}(s)=(W_{\gamma}^{-1}\mathbf{J})(s)$ and
noting the fact that $\mathbf{X}_{0}\in T_{p_{0}}\mathcal{M}$ is
arbitrary, we finally obtain the equation
\begin{equation}
\frac{d^{2}\tilde{\mathbf{J}}(s)}{ds^{2}}+(W_{\gamma}^{-1}R_{\gamma(s)})(\tilde{\mathbf{J}}(s),\mathbf{T}_{0})(\mathbf{T}_{0})=\mathbf{0}\label{eq:geodesic_dev_to_parametric_osc_2}
\end{equation}
The second term on the left hand side of Eq. (\ref{eq:geodesic_dev_to_parametric_osc_2})
can be regarded as a linear transformation of $\tilde{\mathbf{J}}(s)$.
Indeed, if we set
\[
R_{s}\mathbf{X}\,:=(W_{\gamma}^{-1}R_{\gamma(s)})(\mathbf{X},\mathbf{T}_{0})\mathbf{T}_{0},\quad\forall\mathbf{X}\in\ T_{p_{0}}\mathcal{M}
\]
then, for each $s\in\mathbb{R}$, $R_{s}\,:\, T_{p_{0}}\mathcal{M}\mapsto T_{p_{0}}\mathcal{M}$
is a linear operator on $T\mathcal{M}$. Thus we find that the Jacobi
field $\mathbf{J}(s)$ satisfies the geodesic deviation equation,
Eq. (\ref{eq:geodesic_dev_eqn}), if and only if the vector valued
function $\tilde{\mathbf{J}}(\cdot)\in C^{1}(\mathbb{R};\, T_{p_{0}}\mathcal{M})$
satisfies the equation
\begin{equation}
\frac{d^{2}\tilde{\mathbf{J}}(s)}{ds^{2}}+R_{s}\tilde{\mathbf{J}}(s)=\mathbf{0}\label{eq:op_valued_para_osc_eqn}
\end{equation}
where $R_{s}\,:\,\mathbb{R}\mapsto\mathcal{B}(T_{p_{0}}\mathcal{M})$
is an operator valued function defined on $\mathbb{R}$ and taking
values in the space $\mathcal{B}(T_{p_{0}}\mathcal{M})$ of bounded
linear operators on $T_{p_{0}}\mathcal{M}.$ We regard Eq. (\ref{eq:op_valued_para_osc_eqn})
to be an operator valued parametric oscillator equation.

The result in Eq. (\ref{eq:op_valued_para_osc_eqn}) shows rigorously
that the geodesic deviation equation, containing a second order covariant
derivative, is exactly representable, via a unitary transformation,
by a parametric oscillator equation with ordinary second derivative.

\subsection{Dynamical system representation of the geodesic deviation equation\label{sub:dynamical_system_rep}}

We have seen above that solutions $\mathbf{J}(s)$ of the geodesic
deviation equation, Eq. (\ref{eq:geodesic_dev_eqn}), are mapped,
via an application of the map $W_{\gamma}^{-1}$, into solutions $\tilde{\mathbf{J}}(s)$
of the (generalized) parametric oscillator equation, Eq. (\ref{eq:op_valued_para_osc_eqn}).
Moreover, the mapping $W_{\gamma}$ is one-to-one and onto and, hence,
the behavior of solutions of Eq. (\ref{eq:geodesic_dev_eqn}) can
be studied by an analysis of the behavior of the corresponding solutions
of Eq. (\ref{eq:op_valued_para_osc_eqn}). However, if our goal is
the study of the stability properties of the geodesic deviation equation
then Eq. (\ref{eq:op_valued_para_osc_eqn}) presents certain difficulties.
In order to understand these difficulties and the way of overcoming
them and to prepare the setting for the discussion in the next section,
consider the simplest case in which $R_{s}$ is independent of $s$,
i.e., $R_{s}=R_{0}$, $\forall s\in\mathbb{R}$. Explicitly, in this
case we have, for all $\mathbf{X}\in T_{p_{0}}\mathcal{M}$,
\[
R_{s}\mathbf{X}=(W_{\gamma}^{-1}R_{\gamma(s)})(\mathbf{X},\mathbf{T}_{0})(\mathbf{T}_{0})=R_{\gamma(0)}(\mathbf{X},\mathbf{T}_{0})(\mathbf{T}_{0})=R_{0}\mathbf{X}.
\]
By the arbitrariness of $\mathbf{X}$ and $\mathbf{T}_{0}$ (note
that the results of Subsection 2.1 apply to arbitrary geodesics starting
at the point $p_{0}$) this implies that 
\[
R_{p_{0}}=R_{\gamma(0)}=W_{\gamma}^{-1}R_{\gamma(s)}\ \Rightarrow\ R_{\gamma(s)}=(W_{\gamma}R_{p_{0}})(s)=\phi(s)R_{\gamma(0)}
\]
i.e., $R_{s}$ is independent of $s$ if $R_{\gamma(s)}$ is the parallel
transport of $R_{p_{0}}$ along $\gamma$. In this simple case Eq.
(\ref{eq:op_valued_para_osc_eqn}) reduces to
\begin{equation}
\frac{d^{2}\tilde{\mathbf{J}}(s)}{ds^{2}}+R_{0}\tilde{\mathbf{J}}(s)=\mathbf{0}\label{eq:op_valued_osc_eqn-1}
\end{equation}

By the symmetries of the curvature tensor one can prove, using the
Bianchi identity, that $R_{0}$ is a self-adjoint operator on the
real, finite dimensional, Hilbert space $T_{p_{0}}\mathcal{M}$, i.e.,
the matrix reperesenting $R_{0}$ is symmetric. Thus, the eigenvalues
of $R_{0}$ are real. We exclude for the moment the class of operators
having non-trivial kernel and consider an operator $R_{0}$ having
a spectrum $\sigma(R_{0})$ consisting of positive eigenvalues $\omega_{i}^{2}>0$,
$i=1,\ldots,q_{1}$ with corresponding multiplicities $k_{i}$, $i=1,\ldots,,q_{1}$
and negative eigenvalues $-\eta_{j}^{2}<0$, $j=1,\ldots,q_{2}$ with
corrsponding multiplicities $l_{j}$, $j=1,\ldots,q_{2}$ with $\sum_{i=1}^{q_{1}}k_{i}+\sum_{j=1}^{q_{2}}l_{j}=\dim\mathcal{M}=n$.
In order to fix our conventions we set $\omega_{i}>0$, $i=1,\ldots,q_{1}$
in the case of the positive eigenvalues of $R_{0}$ and $\eta_{j}>0$,
$j=1,\ldots,q_{2}$ in the case of the negative eigenvalues of $R_{0}$.
Furthermore, for $i=1,\ldots,q_{1}$, let $\left\{ \hat{\mathbf{v}}_{\omega_{i},r_{i}}\right\} _{r_{i}=1,\ldots,k_{i}}$
be an orthonormal basis for the eigenspace $E_{\omega_{i}}\subseteq T_{p_{0}}\mathcal{M}$
corresponding to the eigenvalue $\omega_{i}^{2}\in\sigma(R_{0})$
and, similarly, for $j=1,\ldots,q_{2}$, let $\left\{ \hat{\mathbf{w}}_{\eta_{j},r_{j}}\right\} _{r_{j}=1,\ldots,l_{j}}$be
an orthonormal basis for the eigenspace $E_{\eta_{j}}\subseteq T_{p_{0}}\mathcal{M}$
corresponding to the eigenvalue $-\eta_{j}^{2}\in\sigma(R_{0})$. 

Consider a positive eigenvalue $\omega_{i}^{2}>0$ of $R_{0}$ and
an arbitrary eigenvector $\mathbf{v}_{\omega_{i}}\in E_{\omega_{i}}$
corresponding to this eigenvalue. We observe that 
\[
\tilde{\mathbf{J}}_{\omega_{i}}(s)=(c_{1}e^{i\omega_{i}s}+c_{2}e^{-i\omega_{i}s})\mathbf{v}_{\omega_{i}}
\]
is an oscillating solution of Eq. (\ref{eq:op_valued_osc_eqn-1}).
Now, if $-\eta_{j}^{2}<0$ is a negative eigenvalue of $R_{0}$ and
$\mathbf{w}_{\eta_{j}}\in E_{\eta_{j}}$ is an eigenvector corresponding
to this eigenvalue, then 
\[
\tilde{\mathbf{J}}_{\eta_{j}}(s)=(c_{1}e^{\eta_{j}s}+c_{2}e^{-\eta_{j}s})\mathbf{\mathbf{w}}_{\eta_{j}}
\]
is a solution of Eq. (\ref{eq:op_valued_osc_eqn-1}). Thus $\tilde{\mathbf{J}}_{\eta_{j}}(s)$
contains in this case both a stable (exponentially decaying) term
and an unstable (exponentially increasing) term corresponding to the
same eigenvector $\mathbf{w}_{\eta_{j}}$ of $R_{0}$ and we cannot
associate $\mathbf{w}_{\eta_{j}}$ uniquely with either stable behavior
or unstable behavior of the corresponding solution $\tilde{\mathbf{J}}_{\eta_{j}}(s)$.
The origin of this property is, of course, is in the fact that Eq.
(\ref{eq:op_valued_para_osc_eqn}) is a second order equation and
that an appropriate initial condition determining a unique solution
of Eq. (\ref{eq:op_valued_para_osc_eqn}) consists of a pair $(\tilde{\mathbf{J}}(0),\,\frac{d\tilde{\mathbf{J}}}{ds}(0)).$
A way of overcoming this difficulty is by constructing a dynamical
system representation incorporating both Eq. (\ref{eq:op_valued_para_osc_eqn})
and the appropriate initial conditions determining a unique solution
of this equation. For this we rewrite Eq. (\ref{eq:op_valued_para_osc_eqn})
in the equivalent form of a non-autonomous, linear dynamical system
\begin{equation}
\frac{d}{ds}\begin{pmatrix}\tilde{\mathbf{J}}(s)\\
\frac{d\tilde{\mathbf{J}}(s)}{ds}
\end{pmatrix}=\begin{pmatrix}0 & I\\
-R_{s} & 0
\end{pmatrix}\begin{pmatrix}\tilde{\mathbf{J}}(s)\\
\frac{d\tilde{\mathbf{J}}(s)}{ds}
\end{pmatrix}=\tilde{R}_{s}\begin{pmatrix}\tilde{\mathbf{J}}(s)\\
\frac{d\tilde{\mathbf{J}}(s)}{ds}
\end{pmatrix}\label{eq:op_valued_parametric_osc_dynamical_sys-1}
\end{equation}
where
\begin{equation}
\tilde{R}_{s}:=\begin{pmatrix}0 & I\\
-R_{s} & 0
\end{pmatrix}.\label{eq:para_osc_dyn_sys_operator_func}
\end{equation}
The dynamical system in Eq. (\ref{eq:op_valued_parametric_osc_dynamical_sys-1})
is defined on the real vector space $T_{p_{0}}\mathcal{M}\oplus T_{p_{0}}\mathcal{M}$.
However, unlike $R_{s}$, the operator $\tilde{R}_{s}$ is not self-adjoint
and, if we lift it to the complexified Hilbert space $\mathcal{H}_{p_{0}}:=\mathbb{C}\otimes(T_{p_{0}}\mathcal{M}\oplus T_{p_{0}}\mathcal{M})$,
then its full spectrum is not a subset of $\mathbb{R}$. Therefore
we lift Eq. (\ref{eq:op_valued_parametric_osc_dynamical_sys-1}) and
consider it to be defined on $\mathcal{H}_{p_{0}}$. Note that the
dynamical system representation in Eq. (\ref{eq:op_valued_parametric_osc_dynamical_sys-1})
corresponds to putting Eq. (\ref{eq:geodesic_dev_eqn}) into the equivalent
form
\begin{equation}
\frac{\nabla}{ds}\begin{pmatrix}\mathbf{J}(s)\\
\frac{\nabla\mathbf{J}(s)}{ds}
\end{pmatrix}=\begin{pmatrix}0 & I\\
-R_{\gamma(s)}(\cdot,\mathbf{T}(s))(\mathbf{T}(s)) & 0
\end{pmatrix}\begin{pmatrix}\mathbf{J}(s)\\
\frac{\nabla\mathbf{J}(s)}{ds}
\end{pmatrix}\label{eq:geodesic_dev_eqn_dynamical_sys-1}
\end{equation}
and applying to Eq. (\ref{eq:geodesic_dev_eqn_dynamical_sys-1}) the
mapping $W_{\gamma}^{-1}$ (more accurately an extension of $W_{\gamma}^{-1}$
to $T_{p}\mathcal{M}\oplus T_{p}\mathcal{M}$). Eq. (\ref{eq:op_valued_parametric_osc_dynamical_sys-1})
is, therefore, a dynamical system representaion of the geodesic deviation
equation, Eq. (\ref{eq:geodesic_dev_eqn}). 

In the simple case that $R_{s}=R_{0}$, $\forall s\in\mathbb{R}$,
Eq. (\ref{eq:op_valued_parametric_osc_dynamical_sys-1}) reduces to
\begin{equation}
\frac{d}{ds}\begin{pmatrix}\tilde{\mathbf{J}}(s)\\
\frac{d\tilde{\mathbf{J}}(s)}{ds}
\end{pmatrix}=\tilde{R}_{0}\begin{pmatrix}\tilde{\mathbf{J}}(s)\\
\frac{d\tilde{\mathbf{J}}(s)}{ds}
\end{pmatrix}\label{eq:op_valued_osc_dynamical_sys}
\end{equation}
with
\begin{equation}
\tilde{R}_{0}:=\begin{pmatrix}0 & I\\
-R_{0} & 0
\end{pmatrix}\label{eq:osc_dyn_sys_generator}
\end{equation}
Now, if $\hat{\mathbf{v}}_{\omega_{i},r_{i}}\in E_{\omega_{i}}$ is
an (basis) eigenvector corresponding to the positive eigenvalue $\omega_{i}^{2}>0$
of $R_{0}$ then the vectors 

\[
\mathbf{u}_{\omega_{i},r_{i}}=\begin{pmatrix}\hat{\mathbf{v}}_{\omega_{i},r_{i}}\\
i\omega_{i}\hat{\mathbf{v}}_{\omega_{i},r_{i}}
\end{pmatrix},\quad\overline{\mathbf{u}}_{\omega_{i},r_{i}}=\begin{pmatrix}\hat{\mathbf{v}}_{\omega_{i},r_{i}}\\
-i\omega_{i}\hat{\mathbf{v}}_{\omega_{i},r_{i}}
\end{pmatrix},\quad\mathbf{u}_{\omega_{i},r_{i}},\overline{\mathbf{u}}_{\omega_{i},r_{i}}\in\mathcal{H}_{p_{0}}
\]
 satisfy
\[
\tilde{R}_{0}\mathbf{u}_{\omega_{i},r_{i}}=i\omega_{i}\mathbf{u}_{\omega_{i},r_{i}},\quad\tilde{R}_{0}\overline{\mathbf{u}}_{\omega_{i},r_{i}}=-i\omega_{i}\overline{\mathbf{u}}_{\omega,r_{i}}
\]
The functions 
\begin{equation}
\tilde{\mathbf{J}}_{i\omega_{i},r_{i}}(s)=e^{i\omega_{i}s}\mathbf{u}_{\omega_{i},r_{i}},\quad\tilde{\mathbf{J}}_{-i\omega_{i},r_{i}}(s)=e^{-i\omega_{i}s}\overline{\mathbf{u}}_{\omega_{i},r_{i}},\quad\forall s\in\mathbb{R}\label{eq:osc_sol_1}
\end{equation}
are then corresponding oscillating solutions of Eq. (\ref{eq:op_valued_parametric_osc_dynamical_sys-1}).

Next consider negative eigenvalues of $R_{0}$. If $\hat{\mathbf{w}}_{\eta_{j},r_{j}}\in E_{\eta_{j}}$
is an (basis) eigenvector corresponding to the negative eigenvalue
$-\eta_{j}^{2}<0$ of $R_{0}$ then the vectors

\[
\mathbf{u}_{\eta_{j},r_{j}}^{+}=\begin{pmatrix}\hat{\mathbf{w}}_{\eta_{j},r_{j}}\\
\eta\hat{\mathbf{w}}_{\eta_{j},r_{j}}
\end{pmatrix},\quad\mathbf{u}_{\eta_{j},r_{j}}^{-}=\begin{pmatrix}\hat{\mathbf{w}}_{\eta_{j},r_{j}}\\
-\eta\hat{\mathbf{w}}_{\eta_{j},r_{j}}
\end{pmatrix},\quad\mathbf{u}_{\eta_{j},r_{j}}^{+},\mathbf{u}_{\eta_{j},r_{j}}^{-}\in\mathcal{H}_{p_{0}}
\]
satisfy
\[
\tilde{R}_{0}\mathbf{u}_{\eta_{j},r_{j}}^{+}=\eta_{j}\mathbf{u}_{\eta_{j},r_{j}}^{+},\quad\tilde{R}_{0}\mathbf{u}_{\eta_{j},r_{j}}^{-}=-\eta_{j}\mathbf{u}_{\eta_{j},r_{j}}^{-}
\]
The functions 
\begin{equation}
\tilde{\mathbf{J}}_{\eta_{j},r_{j}}(s)=e^{\eta_{j}s}\mathbf{u}_{\eta_{j},r_{j}}^{+},\quad\tilde{\mathbf{J}}_{-\eta_{j},r_{j}}(s)=e^{-\eta_{j}s}\mathbf{u}_{\eta_{j},r_{j}}^{-},\quad\forall s\in\mathbb{R}\label{eq:stable_unstable_sol}
\end{equation}
are then, respectively, an unstable solution and a stable solution
of Eq. (\ref{eq:op_valued_osc_dynamical_sys}). We conclude that for
each positive eigenvalue $\omega_{i}^{2}>0$ of $R_{0}$ both $i\omega_{i}$
and $-i\omega_{i}$ are eigenvalues of $\tilde{R}_{0}$ and for each
negative eigenvalue $-\eta_{j}^{2}<0$ of $R_{0}$ both $\eta_{j}$
and $-\eta_{j}$ are eigenvalues of $\tilde{R}_{0}$. Furthermore,
$\tilde{E}_{i\omega_{i}}=span\left\{ \mathbf{u}_{\omega_{i},r_{i}}\right\} _{r_{i}=1,\ldots,k_{i}}$,
$\tilde{E}_{-i\omega_{i}}=span\left\{ \overline{\mathbf{u}}_{\omega_{i},r_{i}}\right\} _{r_{i}=1,\ldots,k_{i}}$,
$\tilde{E}_{\eta_{j}}=span\left\{ \mathbf{u}_{\eta_{j},r_{j}}^{+}\right\} _{r_{j}=1,\ldots,l_{j}}$
and $\tilde{E}_{-\eta_{j}}=span\left\{ \mathbf{u}_{\eta_{j},r_{j}}^{-}\right\} _{r_{j}=1,\ldots,l_{j}}$
are, respectively, $k_{i}$, $k_{i}$, $l_{j}$ and $l_{j}$ dimensional
subspaces of $\mathcal{H}_{p_{0}}$which are eigenspaces, respectively,
for the eigenvalues $i\omega_{i}$, $-i\omega_{i}$, $\eta_{j}$ and
$-\eta_{j}$ of $\tilde{R}_{0}$ . These eigenspaces satisfy 
\[
\mathcal{H}_{p_{0}}=\mathbb{C}\otimes\left(T\mathcal{M}_{p_{0}}\oplus T\mathcal{M}_{p_{0}}\right)=\left(\oplus_{i=1}^{q_{1}}\tilde{E}_{i\omega_{i}}\right)\oplus\left(\oplus_{i=1}^{q_{1}}\tilde{E}_{-i\omega_{i}}\right)\oplus\left(\oplus_{j=1}^{q_{2}}\tilde{E}_{\eta_{j}}\right)\oplus\left(\oplus_{j=1}^{q_{2}}\tilde{E}_{-\eta_{j}}\right).
\]
If we set
\[
\mathcal{H}_{p_{0}}^{c}:=\left(\oplus_{i=1}^{q_{1}}\tilde{E}_{i\omega_{i}}\right)\oplus\left(\oplus_{i=1}^{q_{1}}\tilde{E}_{-i\omega_{i}}\right),\quad\mathcal{H}_{p_{0}}^{u}:=\oplus_{j=1}^{q_{2}}\tilde{E}_{\eta_{j}},\quad\mathcal{H}_{p_{0}}^{s}:=\oplus_{j=1}^{q_{2}}\tilde{E}_{-\eta_{j}}
\]
then according to Eqns. (\ref{eq:osc_sol_1})-(\ref{eq:stable_unstable_sol})
$\mathcal{H}_{p_{0}}^{s}$ is a stable manifold, $\mathcal{H}_{p_{0}}^{u}$
is an unstable manifold and $\mathcal{H}_{p_{0}}^{c}$ is an oscillating
(or central) manifold for the dynamical system in Eq. (\ref{eq:op_valued_osc_dynamical_sys})
and we have a decomposition of $\mathcal{H}_{p_{0}}$ in the form
\begin{equation}
\mathcal{H}_{p_{0}}=\mathcal{H}_{p_{0}}^{c}\oplus\mathcal{H}_{p_{0}}^{s}\oplus\mathcal{H}_{p_{0}}^{u}\label{eq:H_p_zero_decomp}
\end{equation}
We conclude the present section by restating its main result, i.e.,
the fact that the geodesic deviation equation, Eq. (\ref{eq:geodesic_dev_eqn}),
can be represented in terms of the non-autonomous, linear dynamical
system in Eq. (\ref{eq:op_valued_parametric_osc_dynamical_sys-1})
defined on the complex Hilbert space $\mathcal{H}_{p_{0}}=\mathbb{C}\otimes(T_{p_{0}}\mathcal{M}\oplus T_{p_{0}}\mathcal{M})$.
We shall see that the restriction of the evolution of the dynamical
system to $\mathcal{H}_{p_{0}}^{s}$ and $\mathcal{H}_{p_{0}}^{u}$
correspond to semigroups to which the dilation procedure of Sz.-Nagy-Foias
\cite{key-5}, followed by second quantization, may be applied.

\section{Isometric dilation and second quantization of the geodesic deviation
equation\label{sec:isometric_dilation_sec_quant}}

\subsection{Isometric dilation of stable and unstable evolution of the geodesic
deviation equation\label{sub:isometric_dilation_stable_unstable_evol}}

In Sec \ref{sec:parametric_osc_dyn_sys}. we have seen that the geodesic
deviation equation, Eq. (\ref{eq:geodesic_dev_eqn}), corresponding
to the geodesic flow generated by a geometric Hamiltonian of the form
given in Eq. (\ref{eq:geometric_hamiltonian-1}) on a Riemannian manifold
$\mathcal{M}$, can be transformed into an operator valued parametric
oscillator equation, Eq. (\ref{eq:op_valued_para_osc_eqn}), and subsequently
into a linear, non-autonomous, dynamical system, Eq. (\ref{eq:op_valued_parametric_osc_dynamical_sys-1}).
Under the assumption that the operator $\tilde{R}_{s}$ in Eqns. (\ref{eq:op_valued_parametric_osc_dynamical_sys-1})-(\ref{eq:para_osc_dyn_sys_operator_func})
satisfies $\tilde{R}_{s}=\tilde{R}_{0}$, $\forall s\in\mathbb{R}$,
which amounts, as we have seen above, to the simple case where the
curvature tensor $R_{\gamma(s)}$ at the point $\gamma(s)\in\gamma$
is the parallel transport along $\gamma$ of the curvature tensor
$R_{p_{0}}$ at the point $p_{0}\in\gamma$, we obtain the simple
linear autonomous dynamical system in Eq. (\ref{eq:op_valued_osc_dynamical_sys}).
The generator of evolution of this dynamical system is the operator
$\tilde{R}_{0}$, given in Eq. (\ref{eq:osc_dyn_sys_generator}),
i.e., if $\left\{ \tilde{\phi}_{0}(s)\right\} _{s\in\mathbb{R}}$
denotes the evolution of the dynamical system in Eq. (\ref{eq:op_valued_osc_dynamical_sys})
then, given an initial condition $(\tilde{\mathbf{J}}(0),\frac{d\tilde{\mathbf{J}}}{ds}(0))^{T}\in\mathcal{H}_{p_{0}}$,
we have
\[
\left(\begin{array}{c}
\tilde{\mathbf{J}}(s)\\
\frac{d\tilde{\mathbf{J}}}{ds}(s)
\end{array}\right)=\tilde{\phi}_{0}(s)\left(\begin{array}{c}
\tilde{\mathbf{J}}(0)\\
\frac{d\tilde{\mathbf{J}}}{ds}(0)
\end{array}\right)=e^{\tilde{R}_{0}s}\left(\begin{array}{c}
\tilde{\mathbf{J}}(0)\\
\frac{d\tilde{\mathbf{J}}}{ds}(0)
\end{array}\right)
\]
An analysis of the spectrum of $\tilde{R}_{0}$ leads to the decomposition
in Eq. (\ref{eq:H_p_zero_decomp}) of the Hilbert space $\mathcal{H}_{p_{0}}$
into stable, unstable and oscillating (central) subspaces with respect
to forward evolution (i.e., for positive values of $s$). In the present
section our goal is the application of the procedure of dilation,
followed by second quantization, to the stable part and unstable part
of the evolution $\tilde{\phi}_{0}(s)$. As mentioned in Sec. \ref{sec:introduction}
the dilation introduces degrees of freedom corresponding to a dynamical
environment inducing the stability in the case of the stable part
and instability in the case of the unstable part of the evolution
of the dynamical system in Eq. (\ref{eq:op_valued_osc_dynamical_sys}).
The procedure of second quantization exhibits the quantum field associated
with this dynamical environment in the transition to a quantum mechanical
model. Of course, the simple case where $\tilde{R}_{s}=\tilde{R}_{0}$,
$\forall s\in\mathbb{R}$, so convenient for the identification of
the additional degrees of freedom corresponding to the dynamical environment
affecting the stability of the system, is not the generic case and
can be considered as a first order approximation which is a good approximation
only in the case where $\tilde{R}_{s}$ is a function depending very
slowly on the parameter $s$.

At this point we are ready to apply the procedure of dilation to the
stable and unstable parts of the evolution of the dynamical system
in Eq. (\ref{eq:op_valued_osc_dynamical_sys}). We start with the
stable part of this evolution. First note that for $\tau\geq0$ the
stable subspace $\mathcal{H}_{p_{0}}^{s}\subset\mathcal{H}_{p_{0}}$
is invariant under the evolution $\tilde{\phi}_{0}(\tau)$ . For every
$\tau\geq0$ define the operator $\tilde{Z}_{f}(\tau)\,:\,\mathcal{H}_{p_{0}}^{s}\mapsto\mathcal{\mathcal{H}}_{p_{0}}^{s}$
by
\[
\tilde{Z}_{f}(\tau):=\left.\tilde{\phi}_{0}(\tau)\right|_{\mathcal{H}_{p_{0}}^{s}}=\left.e^{\tilde{R}_{0}\tau}\right|_{\mathcal{H}_{p_{0}}^{s}},\quad\tau\geq0
\]
i.e., for $\tau\geq0$, $\tilde{Z}_{f}(\tau)$ is the restriction
of the evolution $\tilde{\phi}_{0}(\tau)$ to the stable subspace
$\mathcal{H}_{p_{0}}^{s}$. By the definitions of $\tilde{Z}_{f}(\tau)$
and $\mathcal{H}_{p_{0}}^{s}$ we have 
\[
\tilde{Z}_{f}(0)=I_{\mathcal{H}_{p_{0}}^{s}}\quad,\tilde{Z}_{f}(\tau_{1})\tilde{Z}_{f}(\tau_{2})=\tilde{Z}_{f}(\tau_{1}+\tau_{2}),\quad\tau,\tau_{2}\geq0,
\]
where $I_{\mathcal{H}_{p_{0}}^{s}}$ is the identity operator on $\mathcal{H}_{p_{0}}^{s}$.
In addition, $\forall\mathbf{v}\in\mathcal{H}_{p_{0}}^{s}$,
\[
\Vert\tilde{Z}_{f}(\tau)\mathbf{v}\Vert_{\mathcal{H}_{p_{0}}}\leq\Vert\mathbf{v}\Vert_{\mathcal{H}_{p_{0}}},\ \tau\geq0,\quad\lim_{\tau\to\infty}\Vert\tilde{Z}_{f}(\tau)\mathbf{v}\Vert_{\mathcal{H}_{p_{0}}}=0\,.
\]
Thus, $\left\{ \tilde{Z}_{f}(\tau)\right\} _{\tau\geq0}$ is a continuous,
contractive, semigroup on $\mathcal{H}_{p_{0}}^{s}$ tending to zero
in the limit $\tau\to\infty$ (in fact, by the finite dimensionality
of the Hilbert space the limit may be taken in the strong, weak or
operator norm sense which are all equivalent in this case).

According to the Sz.-Nagy-Foias theory of contraction operators and
contractive semigroups on Hilbert space \cite{key-5} for such a semigroup
there exists a \emph{minimal isometric dilation}, i.e., there exists
a Hilbert space $\mathcal{R}_{+}^{f}$, an isometric semigroup $\left\{ U_{+}(\tau)\right\} _{\tau\geq0}$
defined on $\mathcal{R}_{+}^{f},$ a subspace $\mathcal{H}_{p_{0},+}^{s}\subset\mathcal{R}_{+}^{f}$
and an isometric isomorphism $V_{+}\,:\,\mathcal{H}_{p_{0}}^{s}\mapsto\mathcal{H}_{p_{0},+}^{s}$
such that
\[
\tilde{Z}_{f}(\tau)=V_{+}^{*}Z_{f}(\tau)V_{+},\quad\tau\geq0,
\]
where

\[
Z_{f}(\tau):=P_{\mathcal{H}_{p_{0},+}^{s}}U_{+}(\tau)P_{\mathcal{H}_{p_{0},+}^{s}},\quad\tau\geq0,
\]
and $P_{\mathcal{H}_{p_{0},+}^{s}}$ is the orthogonal projection
in $\mathcal{R}_{+}^{f}$ on the subspace $\mathcal{H}_{p_{0},+}^{s}$.
Therefore, for $\tau\geq0$, $\tilde{Z}_{f}(\tau)$ is unitarily equivalent
to the projection of $U_{+}(\tau)$ onto the subspace $\mathcal{H}_{p_{0},+}^{s}\subset\mathcal{R}_{+}^{f}$
representing $\mathcal{H}_{p_{0}}^{s}$. The minimality of this isometric
dilation means that $\mathcal{R}_{+}^{f}=\overline{\vee_{\tau\geq0}U_{+}(\tau)\mathcal{H}_{p_{0},+}^{s}}$.
We refer to $\mathcal{R}_{+}^{f}$ as the dilation Hilbert space for
the (minimal) isometric dilation of $\left\{ \tilde{Z}_{f}(\tau)\right\} _{\tau\geq0}$
and to $\left\{ U_{+}(\tau)\right\} _{\tau\geq0}$ as a (minimal)
isometric dilation of the semigroup $\left\{ \tilde{Z}_{f}(\tau)\right\} _{\tau\geq0}$.
The dilation Hilbert space $\mathcal{R}_{+}^{f}$ is naturally decomposed
into two orthogonal subspaces
\[
\mathcal{R}_{+}^{f}=\mathcal{H}_{p_{0},+}^{s}\oplus\mathcal{D}_{+}^{f}
\]
The subspace $\mathcal{D}_{+}^{f}$, generated in the process of construction
of the isometric dilation, corresponds to the dynamical environment
discussed at the beginning of the present section, inducing the stability
of the evolution corresponding to any initial condition in $\mathcal{H}_{p_{0}}^{s}$.

We proceed with a construction of an explicit representation of the
isometric dilation of the semigroup $\left\{ \tilde{Z}_{f}(\tau)\right\} _{\tau\geq0}$
following the procedure described in Appendix A. For this we consider
the Hilbert space $L^{2}(\mathbb{R};\,\mathcal{H}_{p_{0}})\equiv L^{2}(\mathbb{R};\,\mathbb{C}\otimes(T_{p_{0}}\mathcal{M}\oplus T_{p_{0}}\mathcal{M}))$
of Lebesgue square integrable functions defined on the real axis and
taking values in $\mathcal{\mathcal{H}}_{p_{0}}$. We define the inner
product in this space is to be
\[
\langle\tilde{\mathbf{F}},\,\tilde{\mathbf{G}}\rangle_{L^{2}(\mathbb{R};\mathcal{H}_{p_{0}})}:=\intop_{-\infty}^{\infty}\langle\tilde{\mathbf{F}}(s),\,\tilde{\mathbf{G}}(s)\rangle_{\mathcal{H}_{p_{0}}}ds,\qquad\tilde{\mathbf{F}},\tilde{\mathbf{G}}\in L^{2}(\mathbb{R};\mathcal{H}_{p_{0}})
\]
for which the corresponding norm is

\[
\Vert\tilde{\mathbf{F}}\Vert_{L^{2}(\mathbb{R};\mathcal{H}_{p_{0}})}=\left(\intop_{-\infty}^{\infty}\Vert\tilde{\mathbf{F}}(s)\Vert_{\mathcal{H}_{p_{0}}}^{2}ds\right)^{1/2},\qquad\tilde{\mathbf{F}}\in L^{2}(\mathbb{R};\mathcal{H}_{p_{0}}).
\]
Following the decomposition of $\mathcal{H}_{p_{0}}$ in Eq. (\ref{eq:H_p_zero_decomp})
above, we shall be particularly concerned with the two subspaces of
$L^{2}(\mathbb{R};\,\mathcal{H}_{p_{0}})$ corresponding to the stable
subspace $\mathcal{H}_{p_{0}}^{s}$ and unstable subspace $\mathcal{H}_{p_{0}}^{u}$
of $\mathcal{H}_{p_{0}}$, i.e., the function spaces $L^{2}(\mathbb{R};\,\mathcal{H}_{p_{0}}^{s})$
and $L^{2}(\mathbb{R};\,\mathcal{H}_{p_{0}}^{u})$. 

Let $\tilde{B}_{f}=\left.i\tilde{R}_{0}\right|_{\mathcal{H}_{p_{0}}^{s}}$
denote the generator of $\left\{ \tilde{Z}_{f}(\tau)\right\} _{\tau\geq0}$,
i.e., $\tilde{Z}_{f}(\tau)=\left.e^{\tilde{R}_{0}\tau}\right|_{\mathcal{H}_{p_{0}}^{s}}=e^{-i\tilde{B}_{f}\tau}$,
$\tau\geq0$, then there exists a representation of the isometric
dilation of $\left\{ \tilde{Z}_{f}(\tau)\right\} _{\tau\geq0}$, known
as an outgoing representation (see Appendix A), in which the dilation
Hilbert space is $\mathcal{R}_{+}^{f,out}=\mathcal{H}_{p_{0}}^{s,out}\oplus\mathcal{D}_{+}^{f,out}\subset L^{2}(\mathbb{R},\,\mathcal{H}_{p_{0}}^{s})$
where $\mathcal{D}_{+}^{f,out}=L^{2}(\mathbb{R}_{+},\mathcal{H}_{p_{0}}^{s})$
and where $\mathcal{H}_{p_{0}}^{s,out}=\hat{V}_{+}\mathcal{H}_{p_{0}}^{s}\subset L^{2}(\mathbb{R}_{-},\mathcal{H}_{p_{0}}^{s})$
is a unitary embedding of $\mathcal{H}_{p_{0}}^{s}$ into $\mathcal{R}_{+}^{f,out}$
given by
\[
[\hat{V}_{+}\psi](s)=\begin{cases}
(-2\tilde{B}_{f})^{1/2}\tilde{Z}_{f}(-s)\psi, & s\leq0\\
0, & s>0
\end{cases}\,,\quad\psi\in\mathcal{H}_{p_{0}}^{s}
\]
If $\hat{U}(\tau)\,:\, L^{2}(\mathbb{R},\,\mathcal{H}_{p_{0}}^{s})\mapsto L^{2}(\mathbb{R},\,\mathcal{H}_{p_{0}}^{s})$
is translation to the right by $\tau$ units, i.e.,
\[
\left[\hat{U}(\tau)\tilde{\mathbf{F}}\right](t)=\tilde{\mathbf{F}}(t-\tau),\quad\tilde{\mathbf{F}}\in L^{2}(\mathbb{R},\mathcal{H}_{p_{0}}^{s})
\]
then we have (see Appendix A)
\[
\langle\hat{V}_{+}\phi,\,\hat{U}(\tau)\hat{V}_{+}\psi\rangle_{L^{2}(\mathbb{R},\mathcal{H}_{p_{0}}^{s})}=\langle\phi,\tilde{Z}_{f}(\tau)\psi\rangle_{\mathcal{H}_{p_{0}}^{s}},\quad\forall\phi,\psi\in\mathcal{H}_{p_{0}}^{s},\quad\tau\geq0
\]
so that $\hat{U}(\tau)$ is an isometric dilation of $\left\{ \tilde{Z}_{f}(\tau)\right\} _{\tau\geq0}$
on $\mathcal{R}_{+}^{f,out}$.

The dilation of the unstable part of the evolution of the dynamical
system in Eq. (\ref{eq:op_valued_osc_dynamical_sys}) proceeds along
lines similar to the dilation of the stable part of that evolution.
Noting that the unstable subspace $\mathcal{H}_{p_{0}}^{u}\subset\mathcal{H}_{p_{0}}$
is invariant under the evolution $\tilde{\phi}_{0}(\tau)$ for $\tau\leq0$,
we define the operator $\tilde{Z}_{b}(\tau)\,:\,\mathcal{H}_{p_{0}}^{u}\mapsto\mathcal{H}_{p_{0}}^{u}$
by
\[
\tilde{Z}_{b}(\tau)\,:=\left.\tilde{\phi}_{0}(\tau)\right|_{\mathcal{H}_{p_{0}}^{u}}=\left.e^{\tilde{R}_{0}\tau}\right|_{\mathcal{H}_{p_{0}}^{u}},\quad\tau\leq0.
\]
By the definitions of $\tilde{Z}_{b}(\tau)$ and $\mathcal{H}_{p_{0}}^{u}$
we have
\[
\tilde{Z}_{b}(0)=I_{\mathcal{H}_{p_{0}}^{u}}\quad,\tilde{Z}_{b}(\tau_{1})\tilde{Z}_{b}(\tau_{2})=\tilde{Z}_{b}(\tau_{1}+\tau_{2}),\quad\tau_{1},\tau_{2}\leq0,
\]
where $I_{\mathcal{H}_{p_{0}}^{u}}$ is the identity operator on $\mathcal{H}_{p_{0}}^{u}$,
and furthermore, $\forall\mathbf{v}\in\mathcal{H}_{p_{0}}^{u}$,
\[
\Vert\tilde{Z}_{b}(\tau)\mathbf{v}\Vert_{\mathcal{H}_{p_{0}}}\leq\Vert\mathbf{v}\Vert_{\mathcal{H}_{p_{0}}},\ \tau\leq0,\quad\lim_{\tau\to-\infty}\Vert\tilde{Z}_{b}(\tau)\mathbf{v}\Vert_{\mathcal{H}_{p_{0}}}=0\,.
\]
Therefore $\left\{ \tilde{Z}_{b}(\tau)\right\} _{\tau\leq0}$ is a
continuous, contractive, semigroup on $\mathcal{H}_{p_{0}}^{u}$ tending
to zero in the limit $\tau\to-\infty$. The minimal isometric dilation
of this semigroup consists of a dilation Hilbert space $\mathcal{R}_{+}^{b}$,
an isometric evolution semigroup $\left\{ U_{-}(\tau)\right\} _{\tau\leq0}$
defined on $\mathcal{R}_{+}^{b}$, a subspace $\mathcal{H}_{p_{0},+}^{u}\subset\mathcal{R}_{+}^{b}$
and an isometric isomorphism $V_{-}\,:\,\mathcal{H}_{p_{0}}^{u}\mapsto\mathcal{H}_{p_{0},-}^{u}$
such that
\[
\tilde{Z}_{b}(\tau)=V_{-}^{*}Z_{b}(\tau)V_{-},\quad\tau\leq0,
\]
where
\[
Z_{b}(\tau)=P_{\mathcal{H}_{p_{0},+}^{u}}U_{-}(\tau)P_{\mathcal{H}_{p_{0},+}^{u}},\quad\tau\leq0,
\]
and where $P_{\mathcal{H}_{p_{0},+}^{u}}$ is the orthogonal projection
in $\mathcal{R}_{+}^{b}$ on the subspace $\mathcal{H}_{p_{0},+}^{u}$
representing $\mathcal{H}_{p_{0}}^{u}$. The minimality of the isometric
dilation means that $\mathcal{R}_{+}^{b}=\overline{\vee_{\tau\leq0}U_{-}(\tau)\mathcal{H}_{p_{0},-}^{u}}$.
The dilation Hilbert space $\mathcal{R}_{+}^{b}$ is decomposed into
two orthogonal pieces
\[
\mathcal{R}_{+}^{b}=\mathcal{D}_{+}^{b}\oplus\mathcal{H}_{p_{0},+}^{u}
\]
The subspace $\mathcal{D}_{+}^{b}$ ,generated by the procedure of
dilation, corresponds to a dynamical environment inducing the instability
of the evolution (equivalently, the stability of evolution in the
backward direction) corrsponding to any initial condition in $\mathcal{H}_{p_{0}}^{u}$.

Our next step is the construction of an explicit representation of
the isometric dilation of the semigroup $\left\{ \tilde{Z}_{b}(\tau)\right\} _{\tau\leq0}$
following the procedure described in Appendix A. Thus, if $\tilde{B}_{b}=\left.i\tilde{R}_{0}\right|_{\mathcal{H}_{p_{0}}^{u}}$
denotes the generator of $\left\{ \tilde{Z}_{b}(\tau)\right\} _{\tau\leq0}$,
i.e., $\tilde{Z}_{b}(\tau)=\left.e^{\tilde{R}_{0}\tau}\right|_{\mathcal{H}_{p_{0}}^{u}}=e^{-i\tilde{B}_{b}\tau}$,
$\tau\leq0$, then there exists an outgoing representation of the
isometric dilation of $\left\{ \tilde{Z}_{b}(\tau)\right\} _{\tau\leq0}$
where the dilation Hilbert space is $\mathcal{R}_{+}^{b,out}=\mathcal{H}_{p_{0}}^{u,out}\oplus\mathcal{D}_{+}^{b,out}\subset L^{2}(\mathbb{R},\,\mathcal{H}_{p_{0}}^{u})$
where $\mathcal{D}_{+}^{b,out}=L^{2}(\mathbb{R}_{-},\mathcal{H}_{p_{0}}^{u})$
and where $\mathcal{H}_{p_{0}}^{u,out}=\hat{V}_{-}\mathcal{H}_{p_{0}}^{u}\subset L^{2}(\mathbb{R}_{+},\mathcal{H}_{p_{0}}^{u})$
is a unitary embedding of $\mathcal{H}_{p_{0}}^{u}$ into $\mathcal{R}_{+}^{b,out}$
given by
\[
[\hat{V}_{-}\psi](s)=\begin{cases}
(2\tilde{B}_{b})^{1/2}\tilde{Z}_{b}(-s)\psi, & s\geq0\\
0, & s<0
\end{cases}\,.
\]
If $\hat{U}(\tau)\,:\, L^{2}(\mathbb{R},\,\mathcal{H}_{p_{0}}^{u})\mapsto L^{2}(\mathbb{R},\,\mathcal{H}_{p_{0}}^{u})$
is translation to the right by $\tau$ units, i.e., if
\[
\left[\hat{U}(\tau)\tilde{\mathbf{F}}\right](t)=\tilde{\mathbf{F}}(t-\tau),\quad\tilde{\mathbf{F}}\in L^{2}(\mathbb{R},\mathcal{H}_{p_{0}}^{u})
\]
we have (see Appendix A)
\[
\langle\hat{V}_{-}\phi,\,\hat{U}(\tau)\hat{V}_{-}\psi\rangle_{L^{2}(\mathbb{R},\mathcal{H}_{p_{0}}^{u})}=\langle\phi,\tilde{Z}_{b}(\tau)\psi\rangle_{\mathcal{H}_{p_{0}}^{u}},\quad\forall\phi,\psi\in\mathcal{H}_{p_{0}}^{u},\quad\tau\leq0
\]
so that $\hat{U}(\tau)$ is an isometric dilation of $\left\{ \tilde{Z}_{b}(\tau)\right\} _{\tau\leq0}$
on $\mathcal{R}_{+}^{b,out}$. Thus we obtain functional representations
of the isometric dilations of the stable semigroup $\left\{ \tilde{Z}_{f}(\tau)\right\} _{\tau\geq0}$
and the unstable semigroup $\left\{ \tilde{Z}_{b}(\tau)\right\} _{\tau\leq0}$
on subspaces of the function spaces $L^{2}(\mathbb{R},\mathcal{H}_{p_{0}}^{s})$
and $L^{2}(\mathbb{R},\mathcal{H}_{p_{0}}^{u})$ respectively, which,
in turn, are orthogonal subspaces of $L^{2}(\mathbb{R},\mathcal{H}_{p_{0}})$.

\subsection{Isometric dilations on geodesics\label{sub:isometric_dilations_on_geodesics} }

In this subsection we consider Hilbert spaces of vector valued functions
defined along a geodesic $\gamma$ which may carry representations
of the isometric dilation of the stable semigroup $\left\{ \tilde{Z}_{f}(\tau)\right\} _{\tau\geq0}$
and the unstable semigroup $\left\{ \tilde{Z}_{b}(\tau)\right\} _{\tau\leq0}$. 

Let $\gamma$ be a geodesic parametrized by arc length parametrization
as in Eq. (\ref{eq:geodesic_arc_length_para}) so that $T_{\gamma(s)}\mathcal{M}$
is the tangent space to $\mathcal{M}$ at the point $\gamma(s)\in\gamma$.
Let $T\mathcal{M}\oplus T\mathcal{M}$ denote the direct sum of the
tangent bundle of $\mathcal{M}$ with itself. At each point $p\in\mathcal{M}$
the leaf of this bundle is the direct sum $T_{p}\mathcal{M}\oplus T_{p}\mathcal{M}$.
Denote by $T_{\gamma}\mathcal{M}\oplus T_{\gamma}\mathcal{M}$ the
restriction of $T\mathcal{M}\oplus T\mathcal{M}$ to the geodesic
$\gamma$. Denote by $\mathbb{C}\otimes(T\mathcal{M}\oplus T\mathcal{M})$
the complexification of $T\mathcal{M}\oplus T\mathcal{M}$. The leaf
of this complexified bundle at $p\in\mathcal{M}$ is the complex Hilbert
space $\mathbb{C}\otimes(T_{p}\mathcal{M}\oplus T_{p}\mathcal{M})$.
Let $\mathbb{C}\otimes(T_{\gamma}\mathcal{M}\oplus T_{\gamma}\mathcal{M})$
be the restriction of $\mathbb{C}\otimes(T\mathcal{M}\oplus T\mathcal{M})$
to the geodesic $\gamma$. A section of $\mathbb{C}\otimes(T_{\gamma}\mathcal{M}\oplus T_{\gamma}\mathcal{M})$
is a function $\mathbf{F}$ assigning to each point $p\in\gamma$
a vector $\mathbf{F}(p)\in\mathbb{C}\otimes(T_{p}\mathcal{M}\oplus T_{p}\mathcal{M})$.
We fix the parametrization of $\gamma$ to be the arc length parametrization
and consider such a section $\mathbf{F}$ to be a function of the
arc length parameter $s$. Hence we may use the short notation $\mathbf{F}(s)\equiv\mathbf{F}(\gamma(s))$
and consider the section $\mathbf{F}$ to be a function defined on
$\mathbb{R}$. Finally, denote by $L^{2}(\mathbb{R};\,\mathbb{C}\otimes(T_{\gamma}\mathcal{M}\oplus T_{\gamma}\mathcal{M}))$
the Hilbert space of all sections of $\mathbb{C}\otimes(T_{\gamma}\mathcal{M}\oplus T_{\gamma}\mathcal{M})$
which are Lebesgue square integrable with respect to the arc length
parameter. If $\mathbf{G},\mathbf{F}\in L^{2}(\mathbb{R};\,\mathbb{C}\otimes(T_{\gamma}\mathcal{M}\oplus T_{\gamma}\mathcal{M}))$
are two such sections then their inner product is 

\begin{equation*}
\langle\mathbf{F},\,\mathbf{G}\rangle_{L^{2}(\mathbb{R};\mathbb{C}\otimes(T_{\gamma}\mathcal{M}\oplus T_{\gamma}\mathcal{M}))}:=\intop_{-\infty}^{\infty}\langle\mathbf{F}(s),\,\mathbf{G}(s)\rangle_{\mathbb{C}\otimes(T_{\gamma(s)}\mathcal{M}\oplus T_{\gamma(s)}\mathcal{M})}ds
\end{equation*}
and if $\mathbf{F}\in L^{2}(\mathbb{R};\,\mathbb{C}\otimes(T_{\gamma}\mathcal{M}\oplus T_{\gamma}\mathcal{M}))$
then its norm is given by
\[
\Vert\mathbf{F}\Vert_{L^{2}(\mathbb{R};\mathbb{C}\otimes(T_{\gamma}\mathcal{M}\oplus T_{\gamma}\mathcal{M}))}=\left(\intop_{-\infty}^{\infty}\Vert\mathbf{F}(s)\Vert_{\mathbb{C}\otimes(T_{\gamma(s)}\mathcal{M}\oplus T_{\gamma(s)}\mathcal{M})}^{2}\right)^{1/2}
\]

We know from Eq. (\ref{eq:w_gamma_local_unitarity}) that the mapping
$W_{\gamma}$ is locally a unitary mapping of $T_{p_{0}}\mathcal{M}$
onto $T_{\gamma}\mathcal{M}$. We now use this property to extend
this mapping to a unitary map $\hat{W}_{\gamma}=W_{\gamma}\oplus W_{\gamma}\,:\, L^{2}(\mathbb{R};\mathcal{H}_{p_{0}})\mapsto L^{2}(\mathbb{R};\mathbb{C}\otimes(T_{\gamma}\mathcal{M}\oplus T_{\gamma}\mathcal{M}))$,
i.e., for each function $\mathbf{\tilde{F}}=(\tilde{\mathbf{F}}_{1},\,\mathbf{\tilde{F}}_{2})^{T}\in L^{2}(\mathbb{R};\,\mathcal{H}_{p_{0}})$
we define
\[
[\hat{W}_{\gamma}\mathbf{\tilde{F}}](s)=\left(\begin{array}{c}
[W_{\gamma}\mathbf{\tilde{F}}_{1}](s)\\
{}[W_{\gamma}\mathbf{\tilde{F}}_{2}](s)
\end{array}\right)=\left(\begin{array}{c}
\phi(s)\mathbf{\tilde{F}}_{1}(s)\\
\phi(s)\mathbf{\tilde{F}}_{2}(s)
\end{array}\right)
\]
where $\phi(s)$ is parallel transport along $\gamma$. For $\mathbf{\tilde{F}}=(\tilde{\mathbf{F}}_{1},\,\mathbf{\tilde{F}}_{2})^{T}\in L^{2}(\mathbb{R};\,\mathcal{H}_{p_{0}})$
, $\tilde{\mathbf{G}}=(\tilde{\mathbf{G}}_{1},\,\mathbf{\tilde{G}}_{2})^{T}\in L^{2}(\mathbb{R};\,\mathcal{H}_{p_{0}})$
we then have, using Eq. (\ref{eq:w_gamma_local_unitarity}), 
\begin{multline*}
\langle\tilde{\mathbf{F}},\,\tilde{\mathbf{G}}\rangle_{L^{2}(\mathbb{R};\mathcal{H}_{p_{0}})}
=\intop_{-\infty}^{\infty}\langle\tilde{\mathbf{F}}(s),\,\tilde{\mathbf{G}}(s)\rangle_{\mathcal{H}_{p_{0}}}ds=\\
=\intop_{-\infty}^{\infty}\left(\langle\overline{\tilde{\mathbf{F}}_{1}(s)},\,\tilde{\mathbf{G}}_{1}(s)\rangle_{T_{p_{0}}\mathcal{M}}+\langle\overline{\tilde{\mathbf{F}}_{2}(s)},\,\tilde{\mathbf{G}}_{2}(s)\rangle_{T_{p_{0}}\mathcal{M}}\right)ds=\\
=\intop_{-\infty}^{\infty}\left(\langle\overline{[W_{\gamma}\tilde{\mathbf{F}}_{1}](s)},\,[W_{\gamma}\tilde{\mathbf{G}}_{1}](s)\rangle_{T_{\gamma(s)}\mathcal{M}}+\langle\overline{[W_{\gamma}\tilde{\mathbf{F}}_{2}](s)},\,[W_{\gamma}\tilde{\mathbf{G}}_{2}](s)\rangle_{T_{\gamma(s)}\mathcal{M}}\right)ds=\\
=\langle W_{\gamma}\tilde{\mathbf{F}},\, W_{\gamma}\tilde{\mathbf{G}}\rangle_{L^{2}(\mathbb{R};\mathbb{C}\otimes(T_{\gamma}\mathcal{M}\oplus T_{\gamma}\mathcal{M}))},\quad\tilde{\mathbf{F}},\tilde{\mathbf{G}}\in L^{2}(\mathbb{R};\mathcal{H}_{p_{0}})
\end{multline*}
 It is easy to check that $\hat{W}_{\gamma}$ is surjective. Hence,
$\hat{W}_{\gamma}$ is unitary.

We note an important observation associated with the unitary mapping
$\hat{W}_{\gamma}$. Let $\hat{U}(\tau)\,:\, L^{2}(\mathbb{R};\,\mathcal{H}_{p_{0}})\mapsto L^{2}(\mathbb{R};\,\mathcal{H}_{p_{0}})$
be the operator of translation to the right by $\tau$ units, i.e.,

\[
[\hat{U}(\tau)\tilde{\mathbf{F}}](s)=\tilde{\mathbf{F}}(s-\tau),\quad s\in\mathbb{R},\quad\tilde{\mathbf{F}}\in L^{2}(\mathbb{R};\mathcal{H}_{p_{0}})
\]
This operator is unitary on $L^{2}(\mathbb{R};\,\mathcal{H}_{p_{0}})$.
We would like to see how this operator transforms under the unitary
mapping $\hat{W}_{\gamma}$. Denoting $U_{\gamma}(\tau):=\hat{W}_{\gamma}\hat{U}(\tau)\hat{W}_{\gamma}^{-1}$,
we have
\begin{multline*}
[\hat{W}_{\gamma}\hat{U}(\tau)\tilde{\mathbf{F}})](s)=\left(\begin{array}{c}
[W_{\gamma}(\hat{U}(\tau)\mathbf{\tilde{F}}_{1})](s)\\
{}[W_{\gamma}(\hat{U}(\tau)\mathbf{\tilde{F}}_{2})](s)
\end{array}\right)=\left(\begin{array}{c}
\phi(s)(\hat{U}(\tau)\mathbf{\tilde{F}}_{1})(s)\\
\phi(s)(\hat{U}(\tau)\mathbf{\tilde{F}}_{2})(s)
\end{array}\right)=\left(\begin{array}{c}
\phi(s)\mathbf{\tilde{F}}_{1}(s-\tau)\\
\phi(s)\mathbf{\tilde{F}}_{2}(s-\tau)
\end{array}\right)=\\
=\left(\begin{array}{c}
\phi(\tau)\phi(s-\tau)\mathbf{\tilde{F}}_{1}(s-\tau)\\
\phi(\tau)\phi(s-\tau)\mathbf{\tilde{F}}_{2}(s-\tau)
\end{array}\right)=\phi(\tau)\left(\begin{array}{c}
\phi(s-\tau)\mathbf{\tilde{F}}_{1}(s-\tau)\\
\phi(s-\tau)\mathbf{\tilde{F}}_{2}(s-\tau)
\end{array}\right)=\phi(\tau)[\hat{W}_{\gamma}\mathbf{\tilde{F}}](s-\tau)=\\
=[U_{\gamma}(\tau)\hat{W}_{\gamma}\tilde{\mathbf{F}}](s)
\end{multline*}
Thus we have obtained the result that translation to the right by
$\tau$ units on $L^{2}(\mathbb{R};\,\mathcal{H}_{p_{0}})$ is transformed
into parallel transport by $\tau$ units on $L^{2}(\mathbb{R};\mathbb{C}\otimes(T_{\gamma}\mathcal{M}\oplus T_{\gamma}\mathcal{M}))$.
Now consider the representation of the isometric dilation of $\left\{ \tilde{Z}_{f}(\tau)\right\} _{\tau\geq0}$
on the function space $\mathcal{R}_{+}^{f,out}=\mathcal{H}_{p_{0}}^{s,out}\oplus\mathcal{D}_{+}^{f,out}\subset L^{2}(\mathbb{R},\,\mathcal{H}_{p_{0}}^{s})$
introduced in the previous subsection. We apply the mapping $\hat{W}_{\gamma}$
to $\mathcal{R}_{+}^{f,out}$ and set 
\[
\mathcal{R}_{+,\gamma}^{s,out}:=\hat{W}_{\gamma}\mathcal{R}_{+}^{f,out},\qquad\mathcal{H}_{p_{0},\gamma}^{s,out}:=\hat{W}_{\gamma}\mathcal{H}_{p_{0}}^{s,out},\qquad\mathcal{D}_{+,\gamma}^{f,out}:=\hat{W}_{\gamma}\mathcal{D}_{+,\gamma}^{f,out}\,.
\]
Then, by the unitary of $\hat{W}_{\gamma}$, we have
\[
\mathcal{R}_{+,\gamma}^{s,out}=\mathcal{H}_{p_{0},\gamma}^{s,out}\oplus\mathcal{D}_{+,\gamma}^{f,out}\subset L^{2}(\mathbb{R};\mathbb{C}\otimes(T_{\gamma}\mathcal{M}\oplus T_{\gamma}\mathcal{M}))
\]
with
\[
\mathcal{D}_{+,\gamma}^{f,out}=\hat{W}_{\gamma}\mathcal{D}_{+,\gamma}^{f,out}=L^{2}(\mathbb{R}_{+};\mathbb{C}\otimes(T_{\gamma}\mathcal{M}\oplus T_{\gamma}\mathcal{M}))
\]
and where$\mathcal{H}_{p_{0},\gamma}^{s,out}$ is a unitary embedding
of $\mathcal{H}_{p_{0}}^{s}$ into $L^{2}(\mathbb{R};\mathbb{C}\otimes(T_{\gamma}\mathcal{M}\oplus T_{\gamma}\mathcal{M}))$
given explicitly by
\[
[\hat{V}_{+,\gamma}\psi](s)=[\hat{W}_{\gamma}\hat{V}_{+}\psi](s)=\begin{cases}
\phi(s)(-2\tilde{B}_{f})^{1/2}\tilde{Z}_{f}(-s)\psi, & s\leq0\\
0, & s>0
\end{cases}\,,\quad\psi\in\mathcal{H}_{p_{0}}^{s}
\]
where $\hat{V}_{+,\gamma}\,:\,\mathcal{H}_{p}^{s}\mapsto\mathcal{H}_{p_{0},\gamma}^{s,out}$
is defined by $\hat{V}_{+,\gamma}:=\hat{W}_{\gamma}\hat{V}_{+}$ .
We then have

\begin{multline*}
\langle\hat{V}_{+,\gamma}\phi,\, U_{\gamma}(\tau)\hat{V}_{+,\gamma}\psi\rangle_{L^{2}(\mathbb{R};\mathbb{C}\otimes(T_{\gamma}\mathcal{M}\oplus T_{\gamma}\mathcal{M}))}=\langle\hat{W}_{\gamma}\hat{V}_{+}\phi,\, U_{\gamma}(\tau)\hat{W}_{\gamma}\hat{V}_{+}\psi\rangle_{L^{2}(\mathbb{R};\mathbb{C}\otimes(T_{\gamma}\mathcal{M}\oplus T_{\gamma}\mathcal{M}_{\gamma}))}=\\
=\langle\hat{V}_{+}\phi,\,\hat{W}_{\gamma}^{*}U_{\gamma}(\tau)\hat{W}_{\gamma}\hat{V}_{+}\psi\rangle_{L^{2}(\mathbb{R};\mathcal{H}_{p_{0}})}=\langle\hat{V}_{+}\phi,\,\hat{W}_{\gamma}^{-1}U_{\gamma}(\tau)\hat{W}_{\gamma}\hat{V}_{+}\psi\rangle_{L^{2}(\mathbb{R};\mathcal{H}_{p_{0}})}=\\
=\langle\hat{V}_{+}\phi,\,\hat{U}(\tau)\hat{V}_{+}\psi\rangle_{L^{2}(\mathbb{R};\mathcal{H}_{p_{0}})}=\langle\phi,\tilde{Z}_{f}(\tau)\psi\rangle_{\mathcal{H}_{p_{0}}^{s}},\quad\forall\phi,\psi\in\mathcal{H}_{p_{0}}^{s},\quad\tau\geq0
\end{multline*}
so that $U_{\gamma}(\tau)$ is an isometric dilation of the stable
semigroup $\left\{ \tilde{Z}_{f}(\tau)\right\} _{\tau\geq0}$ on $\mathcal{R}_{+,\gamma}^{f,out}\subset L^{2}(\mathbb{R};\mathbb{C}\otimes(T_{\gamma}\mathcal{M}\oplus T_{\gamma}\mathcal{M}))$.
Hence we obtain an isometric dilation of $\left\{ \tilde{Z}_{f}(\tau)\right\} _{\tau\geq0}$
on a Hilbert space of functions defined on the geodesic $\gamma$.
By essentially repeating the same procedure we may obtain an isometric
dilation of the unstable semigroup $\left\{ \tilde{Z}_{b}(\tau)\right\} _{\tau\leq0}$
on a function space $\mathcal{R}_{+,\gamma}^{b,out}\subset L^{2}(\mathbb{R};\mathbb{C}\otimes(T_{\gamma}\mathcal{M}\oplus T_{\gamma}\mathcal{M}))$
defined by $\mathcal{R}_{+,\gamma}^{b,out}:=\hat{W}_{\gamma}\mathcal{R}_{+}^{b,out}$
.

As the geodesic deviation equation is an equation of motion along
a given geodesic $\gamma$, it seems natural to make use of isometric
dilations in function spaces defined over $\gamma$ since they also
utilize motion along the geodesic $\gamma$. However, we emphasize
that, much in the same way that the geodesic deviation equation, Eq.
(\ref{eq:geodesic_dev_eqn}), and its dynamical system representation
in Eq. (\ref{eq:op_valued_parametric_osc_dynamical_sys-1}) are unitarily
equivalent, isometric dilations of $\left\{ \tilde{Z}_{f}(\tau)\right\} _{\tau\geq0}$
and of $\left\{ \tilde{Z}_{b}(\tau)\right\} _{\tau\leq0}$ embedded
in the function space $L^{2}(\mathbb{R};\mathbb{C}\otimes(T_{\gamma}\mathcal{M}\oplus T_{\gamma}\mathcal{M}))$,
defined over the geodesic $\gamma$, and isometric dilations embedded
in the function space $L^{2}(\mathbb{R},\mathcal{H}_{p_{0}})$ of
$\mathcal{H}_{p_{0}}$ valued functions are completly equivalent by
the unitarity of the mapping $\hat{W}_{\gamma}$ and there is no fundamental
reason to favor one of these representations over the other. Athough
the isometric dilation in the function space defined over the geodesic
$\gamma$ is more natural and conceptually important, for the sake
of simplicity the procedure of second quantization of the isometric
dilation spaces is applied in the next subsection within the $L^{2}(\mathbb{R},\mathcal{H}_{p_{0}})$
setting.

\subsection{Second quantization\label{sub:second_quantization}}

Consider the isometric dilation of the stable semigroup $\left\{ \tilde{Z}_{f}(\tau)\right\} _{\tau\geq0}$
constructed in the function space $\mathcal{R}_{+}^{f,out}\subset L^{2}(\mathbb{R},\mathcal{H}_{p_{0}}^{s})$
in Subsection 3.1 above. The subspace $\mathcal{H}_{p_{0}}^{s,out}\subset\mathcal{R}_{+}^{f,out}$
is unitarily equivalent to $\mathcal{H}_{p_{0}}^{s}$ via the mapping
$\hat{V}_{+}$ and the restriction$\left\{ Z_{f}(\tau)\right\} _{\tau\geq0}$,
$Z_{f}(\tau)=P_{\mathcal{H}_{p_{0}}^{s,out}}\hat{U}(\tau)P_{\mathcal{H}_{p_{0}}^{s,out}}$,
of the isometric evolution $\left\{ \hat{U}(\tau)\right\} _{\tau\geq0}$
in $\mathcal{R}_{+}^{f,out}$ to $\mathcal{H}_{p_{0}}^{s,out}$ is
unitarily equivalent to $\left\{ \tilde{Z}_{f}(\tau)\right\} _{\tau\geq0}$
which in turn, as we recall, is a first order approximation to the
stable part of the forward evolution of the geodesic deviation equation,
Eq. (\ref{eq:geodesic_dev_eqn}). As opposed to $\mathcal{H}_{p_{0}}^{s,out}$
the subspace $\mathcal{D}_{+}^{f,out}$ is generated by the dilation
procedure and represents new degrees of freedom which are not part
of the original system. In order to understand the meaning of these
new degrees of freedom and the way they influence the evolution generated
by the geodesic deviation equation, we apply a procedure of second
quantization, identify a quantum field associated with $\mathcal{D}_{+}^{f,out}$
and observe how the interaction of this field with the system induces
the evolution of the stable semigroup $\left\{ \tilde{Z}_{f}(\tau)\right\} _{\tau\geq0}$.
More specifically, we apply coherent state second quantization to
the isometric dilation Hilbert space $\mathcal{R}_{+}^{f,out}=\mathcal{H}_{p_{0}}^{s,out}\oplus\mathcal{D}_{+}^{f,out}$.
For this we define a positive kernel $K(\cdot.\cdot)\,:\,\mathcal{R}_{+}^{f,out}\times\mathcal{R}_{+}^{f,out}\mapsto\mathbb{C}$
by
\[
K(u,v)=e^{\langle u,v\rangle_{\mathcal{R}_{+}^{f,out}}}=e^{\langle u,v\rangle_{L^{2}(\mathbb{R},H_{p_{0}}^{s})}},\quad u,v\in\mathcal{R}_{+}^{f,out}
\]
and perform a Kolmogorov dilation (see, for example, \cite{key-17})
of $\mathcal{R}_{+}^{f,out}$ with respect to $K(\cdot,\cdot)$. The
procedure of Kolomogorov dilation introduces a symmetric Fock space
$\Gamma_{s}(\mathcal{R}_{+}^{f,out})=\sum_{n=0}^{\infty}\oplus\left[\left(\mathcal{R}_{+}^{f,out}\right)^{\otimes^{n}}\right]_{sym}$
such that to every state $u\in\mathcal{R}_{+}^{f,out}$ there is assigned
an exponential vector (which is identical to a coherent state upto
a normalization factor) $e(u)\in\Gamma_{s}(\mathcal{R}_{+}^{f,out})$
\[
e(u)=\sum_{n=0}^{\infty}\frac{1}{\sqrt{n!}}u^{\otimes^{n}}
\]
and we have
\[
\langle e(u),\, e(v)\rangle_{\Gamma_{s}(\mathcal{R}_{+}^{f,out})}=\sum_{n=0}^{\infty}\frac{1}{n!}\langle u,v\rangle_{\mathcal{R}_{+}^{f,out}}^{n}=e^{\langle u,v\rangle_{\mathcal{R}_{+}^{f,out}}}=K(u,v),\quad u,v\in\mathcal{R}_{+}^{f,out}
\]
We define a representation of the canonical commutation relations
(CCR) on $\Gamma_{s}(\mathcal{R}_{+}^{f,out})$ through the algebra
of Weyl operators (see, for example, \cite{key-6}) $W(u,\, U)\,:\,\Gamma_{s}(\mathcal{R}_{+}^{f,out})\mapsto\Gamma_{s}(\mathcal{R}_{+}^{f,out})$,
with $u\in\mathcal{R}_{+}^{f,out}$ and $U\in\mathscr{U}\left(\mathcal{R}_{+}^{f,out}\right)$,
where $\mathscr{U}\left(\mathcal{R}_{+}^{f,out}\right)$ is the group
of unitary operators on $\mathcal{R}_{+}^{f,out}$. The action of
a Weyl operator on exponential vectors is given by
\begin{equation}
W(u,U)e(v)=e^{-\frac{1}{2}\Vert u\Vert_{\mathcal{R}_{+}^{f,out}}^{2}+\langle u,Uv\rangle_{\mathcal{R}_{+}^{f,out}}}e(Uv+u)\label{eq:weyl_op_def}
\end{equation}
The Weyl operators are unitary operators on $\Gamma_{s}(\mathcal{R}_{+}^{f,out})$.
The composition rule for Weyl operators is given by (see, for example,
\cite{key-6})
\begin{equation}
W(u_{2},U_{2})W(u_{1},U_{1})=e^{-\text{Im}\langle u_{2},U_{2}u_{1}\rangle}W(U_{2}u_{1}+u_{2},U_{2}U_{1})\label{eq:weyl_op_composition}
\end{equation}

Considering the special cases
\[
W(u)\,:=W(u,I_{\mathcal{R}_{+}^{f,out}}),\qquad\Gamma(U)\,:=W(0,U)
\]
and using the composition rule of Weyl operators from Eq. (\ref{eq:weyl_op_composition})
we obtain
\begin{equation}
W(u)W(v)=e^{-i\text{Im}\langle u,v\rangle}W(u+v)\label{eq:weyl_properties_1}
\end{equation}
\begin{equation}
W(u)W(v)=e^{-i2\text{Im}\langle u,v\rangle}W(v)W(u)\label{eq:weyl_properties_2}
\end{equation}
\begin{equation}
\Gamma(U_{2})\Gamma(U_{1})=\Gamma(U_{2}U_{1})\label{eq:weyl_properties_3}
\end{equation}
\begin{equation}
\Gamma(U)W(u)\Gamma^{-1}(U)=W(Uu)\label{eq:weyl_properties_4}
\end{equation}
\begin{equation}
W(su)W(tu)=W((s+t)u),\quad s,t\in\mathbb{R}\label{eq:weyl_properties_5}
\end{equation}
Basic observables on $\Gamma_{s}(\mathcal{R}_{+}^{f,out})$ are defined
through one parameter subgroups in the algebra of Weyl operators.
Thus, if $u$ is an element of $\mathcal{R}_{+}^{f,out}$ , Eq. (\ref{eq:weyl_properties_5})
implies that $\left\{ W(tu)\right\} _{t\in\mathbb{R}}$ is a continuous,
one parameter unitary group on $\Gamma_{s}(\mathcal{R}_{+}^{f,out})$.
This group has a self-adjoint generator $p(u)$ so that
\[
W(tu)=e^{-ip(u)t}
\]
and $p(u)$ is a basic observable on the Fock space $\Gamma_{s}(\mathcal{R}_{+}^{f,out})$.
Setting
\[
q(u)\,:=-p(iu)
\]
one defines the operators
\[
a(u)\,:=\frac{1}{2}\left(q(u)+ip(u)\right),\qquad a^{\dagger}(u)\,:=\frac{1}{2}\left(q(u)-ip(u)\right)
\]
which act, respectively, as annihilation and creation operators on
$\Gamma_{s}(\mathcal{R}_{+}^{f,out})$. Another type of basic observables
on $\Gamma_{s}(\mathcal{R}_{+}^{f,out})$ follows from the fact that
if $\left\{ U(t)\right\} _{t\in\mathbb{R}}$ , $U(t)=e^{-iHt}$ is
a continuous, one parameter unitary group on $\mathcal{R}_{+}^{f,out}$
with a self-adjoint generator $H$, then Eq. (\ref{eq:weyl_properties_3})
implies that $\left\{ \Gamma(U(t))\right\} _{t\in\mathbb{R}}=\left\{ \Gamma(e^{-iHt})\right\} _{t\in\mathbb{R}}$
is a continuous, one parameter unitary group defined on $\Gamma_{s}(\mathcal{R}_{+}^{f,out})$
. We denote the self-adjoint generator of this group by $\lambda(H)$
so that
\[
\Gamma(e^{-iHt})=e^{-i\lambda(H)t}
\]
 and $\lambda(H)$ is a basic observable on $\Gamma_{s}(\mathcal{R}_{+}^{f,out})$.
One may think about $\lambda(H)$ as a lifting of the observable $H$
from $\mathcal{R}_{+}^{f,out}$ to the Fock space $\Gamma_{s}(\mathcal{R}_{+}^{f,out})$.
Amongst the observables of the later type we shall consider in particular
those associated with orthogonal projections in $\mathcal{R}_{+}^{f,out}$.
Thus, if $P\,:\,\mathcal{R}_{+}^{f,out}\mapsto\mathcal{R}_{+}^{f,out}$
is an orthogonal projection then $\lambda(P)$ is the corresponding
observable on $\Gamma_{s}(\mathcal{R}_{+}^{f,out})$. It can be shown
that $\lambda(P)$ has a natural interpretation as the observable
that counts the number of quanta in a state $f\in\Gamma_{s}(\mathcal{R}_{+}^{f,out})$
for which the question defined by the projection $P$ (i.e., the question
of whether they belong to the range of $P$) is answered in the affirmative
(${\bf Ref}$.). In other words, if $P^{\perp}:=I_{\mathcal{R}_{+}^{f,out}}-P$,
so that $\mathcal{R}_{+}^{f,out}=\left(P\mathcal{R}_{+}^{f,out}\right)\oplus\left(P^{\perp}\mathcal{R}_{+}^{f,out}\right)$
and hence
\[
\Gamma_{s}(\mathcal{R}_{+}^{f,out})=\Gamma_{s}\left(P\mathcal{R}_{+}^{f,out}\right)\otimes\Gamma_{s}\left(P^{\perp}\mathcal{R}_{+}^{f,out}\right)\,,
\]
then $\lambda(P)$ is the observable counting the number of quanta
in $\Gamma_{s}\left(P\mathcal{R}_{+}^{f,out}\right)$ and $\lambda(P^{\perp})$
is the observable counting the number of quanta in $\Gamma_{s}\left(P^{\perp}\mathcal{R}_{+}^{f,out}\right)$. 

Now, note that by the orthognal direct sum decomposition $\mathcal{R}_{+}^{f,out}=\mathcal{H}_{p_{0}}^{s,out}\oplus\mathcal{D}_{+}^{f,out}$
we have a decomposition of the Fock space $\Gamma_{s}(\mathcal{R}_{+}^{f,out})$
into a tensor product 
\[
\Gamma_{s}(\mathcal{R}_{+}^{f,out})=\Gamma_{s}(\mathcal{H}_{p_{0}}^{s,out})\otimes\Gamma_{s}(\mathcal{D}_{+}^{f,out})
\]
let $P_{+}$ be the orthogonal projection in $L^{2}(\mathbb{R},\mathcal{H}_{p_{0}})$
on the closed subspace $L^{2}(\mathbb{R}_{+},\mathcal{H}_{p_{0}})$
and let $P_{+}^{\perp}:=I_{L^{2}(\mathbb{R},\mathcal{H}_{p_{0}})}-P_{+}$,
so that $P_{+}^{\perp}$ is the orthogonal projection in $L^{2}(\mathbb{R},\mathcal{H}_{p_{0}})$
on the closed subspace $L^{2}(\mathbb{R}_{-},\mathcal{H}_{p_{0}})$.
Since $\mathcal{R}_{+}^{f,out}=\mathcal{H}_{p_{0}}^{s,out}\oplus\mathcal{D}_{+}^{f,out}$
with $D_{+}^{f,out}=L^{2}(\mathbb{R}_{+},\mathcal{H}_{p_{0}})$ and
$\mathcal{H}_{p_{0}}^{s,out}\subset L^{2}(\mathbb{R}_{-},\mathcal{H}_{p_{0}}^{s})$
we find that $\tilde{P}_{+}:=\left.P_{+}\right|_{\mathcal{R}_{+}^{f,out}}$
is the orthogonal projection in $\mathcal{R}_{+}^{f,out}$ on $\mathcal{D}_{+}^{f,out}$
and $\tilde{P}_{+}^{\perp}:=\left.P_{+}^{\perp}\right|_{\mathcal{R}_{+}^{f,out}}$
is the orthogonal projection in $\mathcal{R}_{+}^{f,out}$ on $\mathcal{H}_{p_{0}}^{s,out}$
so that, in particular, we have $\tilde{P}_{+}+\tilde{P}_{+}^{\perp}=I_{\mathcal{R}_{+}^{f,out}}$
and
\[
\Gamma_{s}(\mathcal{R}_{+}^{f,out})=\Gamma_{s}(\mathcal{H}_{p_{0}}^{s,out})\otimes\Gamma_{s}(\mathcal{D}_{+}^{f,out})=\Gamma_{s}\left(\tilde{P}_{+}\mathcal{R}_{+}^{f,out}\right)\otimes\Gamma_{s}\left(\tilde{P}_{+}^{\perp}\mathcal{R}_{+}^{f,out}\right)
\]
The second quantization $\lambda(\tilde{P}_{+})$ and $\lambda(\tilde{P}_{+}^{\perp})$
are then observables on $\Gamma_{s}(\mathcal{R}_{+}^{f,out})$ counting,
respectively, the number of quanta in $\Gamma_{s}(\mathcal{D}_{+}^{f,out})$
and $\Gamma_{s}(\mathcal{H}_{p_{0}}^{s,out})$. Note also that we
have
\begin{multline*}
e^{-i\lambda(\tilde{P}_{+})t}e^{-i\lambda(\tilde{P}_{+}^{\perp})t}=\Gamma(e^{-i\tilde{P}_{+}t})\Gamma(e^{-i\tilde{P}_{+}^{\perp}t})=\Gamma(e^{-i\tilde{P}_{+}t}e^{-i\tilde{P}_{+}^{\perp}t})=\\
=\Gamma(e^{-i(\tilde{P}_{+}+\tilde{P}_{+}^{\perp})t})=\Gamma(e^{-iI_{\mathcal{R}_{+}^{f,out}}t})=e^{-i\lambda(I_{\mathcal{R}_{+}^{f,out}})t}
\end{multline*}
from which we get that 
\[
\lambda(\tilde{P}_{+})+\lambda(\tilde{P}_{+}^{\perp})=\lambda(I_{\mathcal{R}_{+}^{f,out}})
\]
where $\lambda(I_{\mathcal{R}_{+}^{f,out}})$ is the observable counting
the total number of quanta in $\Gamma_{s}(\mathcal{R}_{+}^{f,out})$.

Our next step following the identification of the observables counting
the number of quanta in $\Gamma_{s}(\mathcal{D}_{+}^{f,out})$ and
$\Gamma_{s}(\mathcal{H}_{p_{0}}^{s,out})$ is the lifting of the isometric
evolution $\left\{ \hat{U}(\tau)\right\} _{\tau\geq0}$ from $\mathcal{R}_{+}^{f,out}$
to the Fock space $\Gamma_{s}(\mathcal{R}_{+}^{f,out})$ and the analysis
of the effect of this evolution in $\Gamma_{s}(\mathcal{D}_{+}^{f,out})$
and $\Gamma_{s}(\mathcal{H}_{p_{0}}^{s,out})$. For a general contraction
operator $C\,:\,\mathcal{R}_{+}^{f,out}\mapsto\mathcal{R}_{+}^{f,out}$
the lifting of the action of $C$ into the Weyl algera is denoted
by $\Gamma_{0}(C)$ and defined by

\[
\Gamma_{0}(C)W(u)=e^{\frac{1}{2}\left(\Vert Cu\Vert_{\mathcal{R}_{+}^{f,out}}^{2}-\Vert u\Vert_{\mathcal{R}_{+}^{f,out}}^{2}\right)}W(Cu)
\]
Note that if $C_{1}$ and $C_{2}$ are contractions we have
\begin{multline*}
\Gamma_{0}(C_{2})\Gamma_{0}(C_{1})W(u)=e^{\frac{1}{2}\left(\Vert C_{1}u\Vert_{\mathcal{R}_{+}^{f,out}}^{2}-\Vert u\Vert_{\mathcal{R}_{+}^{f,out}}^{2}\right)}\Gamma_{0}(C_{2})W(C_{1}u)=\\
=e^{\frac{1}{2}\left(\Vert C_{1}u\Vert_{\mathcal{R}_{+}^{f,out}}^{2}-\Vert u\Vert_{\mathcal{R}_{+}^{f,out}}^{2}\right)}e^{\frac{1}{2}\left(\Vert C_{2}C_{1}u\Vert_{\mathcal{R}_{+}^{f,out}}^{2}-\Vert C_{1}u\Vert_{\mathcal{R}_{+}^{f,out}}^{2}\right)}W(C_{2}C_{1}u)=\\
=e^{\frac{1}{2}\left(\Vert C_{2}C_{1}u\Vert_{\mathcal{R}_{+}^{f,out}}^{2}-\Vert u\Vert_{\mathcal{R}_{+}^{f,out}}^{2}\right)}W(C_{2}C_{1}u)=\Gamma_{0}(C_{2}C_{1})W(u)
\end{multline*}
from which we obtain a composition rule $\Gamma_{0}(C_{2})\Gamma_{0}(C_{1})=\Gamma_{0}(C_{2}C_{1})$,
extending the composition rule in Eq. (\ref{eq:weyl_properties_3})
above. Note also that, since an exponential vector $e(v)$ is given
by the action of a Weyl operator on the vacuum state $e(0)$ by
\[
e(v)=e^{\frac{1}{2}\Vert v\Vert_{\mathcal{R}_{+}^{f,out}}^{2}}W(v)e(0)
\]
 then we may extend the action of $\Gamma_{0}(C)$ to exponential
vectors via
\begin{multline*}
\Gamma_{0}(C)e(v)\,:=e^{\frac{1}{2}\Vert v\Vert_{\mathcal{R}_{+}^{f,out}}^{2}}[\Gamma_{0}(C)W(v)]e(0)=e^{\frac{1}{2}\Vert v\Vert_{\mathcal{R}_{+}^{f,out}}^{2}}e^{\frac{1}{2}\left(\Vert Cv\Vert_{\mathcal{R}_{+}^{f,out}}^{2}-\Vert v\Vert_{\mathcal{R}_{+}^{f,out}}^{2}\right)}W(Cv)e(0)=\\
=e^{\frac{1}{2}\Vert Cv\Vert_{\mathcal{R}_{+}^{f,out}}^{2}}e^{-\frac{1}{2}\Vert Cv\Vert_{\mathcal{R}_{+}^{f,out}}^{2}}e(Cv)=e(Cv)
\end{multline*}
If we now take for the contraction operators the elements $\hat{U}(\tau)$
of the isometric evolution $\left\{ \hat{U}(\tau)\right\} _{\tau\geq0}$
we obtain 
\[
\Gamma_{0}(\hat{U}(\tau_{2})\Gamma_{0}(\hat{U}(\tau_{1}))=\Gamma_{0}(\hat{U}(\tau_{2})\hat{U}(\tau_{1}))=\Gamma_{0}(\hat{U}(\tau_{2}+\tau_{1})),\quad\tau_{2},\tau_{1}\geq0
\]
and
\[
\Gamma_{0}(\hat{U}(\tau))e(v)=e(\hat{U}(\tau)v),\quad\tau\geq0,\ v\in\mathcal{R}_{+}^{f,out}
\]
 This defines a lifting of the evolution $\left\{ \hat{U}(\tau)\right\} _{\tau\geq0}$
into the Weyl algebra and, hence, into $\Gamma_{s}(\mathcal{R}_{+}^{f,out})$.

Given an eigenvector $\mathbf{u}_{\eta_{j},r_{j}}^{-}\in\mathcal{H}_{p_{0}}^{s}$
of the stable semigroup $\left\{ \tilde{Z}_{f}(\tau)\right\} _{\tau\geq0}$
satisfying

\[
\tilde{Z}_{f}(\tau)\mathbf{u}_{\eta_{j},r_{j}}^{-}=e^{-\eta_{j}\tau}\mathbf{u}_{\eta_{j},r_{j}}^{-},\quad\tau\geq0,
\]
we apply the unitary mapping $\hat{V}_{+}\,:\,\mathcal{H}_{p_{0}}^{s}\mapsto\mathcal{H}_{p_{0}}^{s,out}$
embedding $\mathcal{H}_{p_{0}}^{s}$ and the semigroup $\left\{ \tilde{Z}_{f}(\tau)\right\} _{\tau\geq0}$
into $\mathcal{R}_{+}^{f,out}$. Thus, if 
\[
Z_{f}(\tau)=\hat{V}_{+}^{*}\tilde{Z}_{f}(\tau)\hat{V}_{+}=P_{\mathcal{H}_{p_{0}}^{s,out}}\hat{U}(\tau)P_{\mathcal{H}_{p_{0}}^{s,out}}=\tilde{P}_{+}^{\perp}\hat{U}(\tau)\tilde{P}_{+}^{\perp},\quad\tau\geq0
\]
then
\[
Z_{f}(\tau)\hat{V}_{+}\mathbf{u}_{\eta_{j},r_{j}}^{-}=e^{-\eta_{j}\tau}\hat{V}_{+}\mathbf{u}_{\eta_{j},r_{j}}^{-},\quad\tau\geq0\,.
\]
If we now apply second quantiztion then $\hat{V}_{+}\mathbf{u}_{\eta_{j},r_{j}}^{-}$
is mapped into an exponential vector $e(\hat{V}_{+}\mathbf{u}_{\eta_{j},r_{j}}^{-})\in\Gamma_{s}(\mathcal{R}_{+}^{f,out})$
and the isometric dilation evolution $\left\{ \hat{U}(\tau)\right\} _{\tau\geq0}$
is lifted into the evolution $\left\{ \Gamma_{0}(\hat{U}(\tau))\right\} _{\tau\geq0}$
in $\Gamma_{s}(\mathcal{R}_{+}^{f,out})$. For any observable $A$
on $\Gamma_{s}(\mathcal{R}_{+}^{f,out})$ the expectation value of
$A$ in the evolved state $\Gamma_{0}(\hat{U}(\tau))e(\hat{V}_{+}\mathbf{u}_{\eta_{j},r_{j}}^{-})$
is, of course, given by 
\begin{multline*}
\langle\Gamma_{0}(\hat{U}(\tau))e(\hat{V}_{+}\mathbf{u}_{\eta_{j},r_{j}}^{-}),\, A\Gamma_{0}(\hat{U}(\tau))e(\hat{V}_{+}\mathbf{u}_{\eta_{j},r_{j}}^{-})\rangle_{\Gamma_{s}(\mathcal{R}_{+}^{f,out})}=\\
=\langle e(\hat{U}(\tau)\hat{V}_{+}\mathbf{u}_{\eta_{j},r_{j}}^{-}),\, Ae(\hat{U}(\tau)\hat{V}_{+}\mathbf{u}_{\eta_{j},r_{j}}^{-})\rangle_{\Gamma_{s}(\mathcal{R}_{+}^{f,out})}
\end{multline*}
For $0<s<1$ define on $\Gamma_{s}(\mathcal{H}_{p_{0}}^{s})$ the
operator $s^{\lambda(\tilde{P}_{+}^{\perp})}$ and recall that $\lambda(\tilde{P}_{+}^{\perp})$
is the operator counting the number of quanta in $\Gamma_{s}(\mathcal{H}_{p_{0}}^{s,out})$.
Calculating the above expectation value with $A=s^{\lambda(\tilde{P}_{+}^{\perp})}$
we obtain
\begin{multline*}
\langle\Gamma_{0}(\hat{U}(\tau))e(\hat{V}_{+}\mathbf{u}_{\eta_{j},r_{j}}^{-}),\, s^{\lambda(\tilde{P}_{+}^{\perp})}\Gamma_{0}(\hat{U}(\tau))e(\hat{V}_{+}\mathbf{u}_{\eta_{j},r_{j}}^{-})\rangle_{\Gamma_{s}(\mathcal{R}_{+}^{f,out})}=\\
=\langle e(\hat{U}(\tau)\hat{V}_{+}\mathbf{u}_{\eta_{j},r_{j}}^{-}),\, e(s^{\tilde{P}_{+}^{\perp}}\hat{U}(\tau)\hat{V}_{+}\mathbf{u}_{\eta_{j},r_{j}}^{-})\rangle_{\Gamma_{s}(\mathcal{R}_{+}^{f,out})}=\\
=\langle e(\hat{U}(\tau)\hat{V}_{+}\mathbf{u}_{\eta_{j},r_{j}}^{-}),\, e(e^{(\ln s)\tilde{P}_{+}^{\perp}}\hat{U}(\tau)\hat{V}_{+}\mathbf{u}_{\eta_{j},r_{j}}^{-})\rangle_{\Gamma_{s}(\mathcal{R}_{+}^{f,out})}=\\
=\langle e(\hat{U}(\tau)\hat{V}_{+}\mathbf{u}_{\eta_{j},r_{j}}^{-}),\, e([e^{(\ln s)}\tilde{P}_{+}^{\perp}+\tilde{P}_{+}]\hat{U}(\tau)\hat{V}_{+}\mathbf{u}_{\eta_{j},r_{j}}^{-})\rangle_{\Gamma_{s}(\mathcal{R}_{+}^{f,out})}=\\
=\langle e(\hat{U}(\tau)\hat{V}_{+}\mathbf{u}_{\eta_{j},r_{j}}^{-}),\, e([(s-1)\tilde{P}_{+}^{\perp}+I_{\mathcal{R}_{+}^{f,out}}]\hat{U}(\tau)\hat{V}_{+}\mathbf{u}_{\eta_{j},r_{j}}^{-})\rangle_{\Gamma_{s}(\mathcal{R}_{+}^{f,out})}=\\
=e^{\langle\hat{U}(\tau)\hat{V}_{+}\mathbf{u}_{\eta_{j},r_{j}}^{-},\,[(s-1)\tilde{P}_{+}^{\perp}+I_{\mathcal{R}_{+}^{f,out}}]\hat{U}(\tau)\hat{V}_{+}\mathbf{u}_{\eta_{j},r_{j}}^{-}\rangle_{\mathcal{R}_{+}^{f,out}}}=\\
=e^{\langle\hat{U}(\tau)\hat{V}_{+}\mathbf{u}_{\eta_{j},r_{j}}^{-},\,(s-1)\tilde{P}_{+}^{\perp}\hat{U}(\tau)\hat{V}_{+}\mathbf{u}_{\eta_{j},r_{j}}^{-}\rangle_{\mathcal{R}_{+}^{f,out}}+\langle\hat{V}_{+}\mathbf{u}_{\eta_{j},r_{j}}^{-},\hat{V}_{+}\mathbf{u}_{\eta_{j},r_{j}}^{-}\rangle_{\mathcal{R}_{+}^{f,out}}}=\\
=e^{(s-1)\Vert\tilde{P}_{+}^{\perp}\hat{U}(\tau)\tilde{P}_{+}^{\perp}\hat{V}_{+}\mathbf{u}_{\eta_{j},r_{j}}^{-}\Vert_{\mathcal{R}_{+}^{f,out}}^{2}+\Vert\hat{V}_{+}\mathbf{u}_{\eta_{j},r_{j}}^{-}\Vert_{\mathcal{R}_{+}^{f,out}}^{2}}=\\
=e^{(s-1)\Vert Z_{f}(\tau)\hat{V}_{+}\mathbf{u}_{\eta_{j},r_{j}}^{-}\Vert_{\mathcal{R}_{+}^{f,out}}^{2}+\Vert\hat{V}_{+}\mathbf{u}_{\eta_{j},r_{j}}^{-}\Vert_{\mathcal{R}_{+}^{f,out}}^{2}}=e^{(s-1)e^{-2\eta_{j}\tau}\Vert\hat{V}_{+}\mathbf{u}_{\eta_{j},r_{j}}^{-}\Vert_{\mathcal{R}_{+}^{f,out}}^{2}+\Vert\hat{V}_{+}\mathbf{u}_{\eta_{j},r_{j}}^{-}\Vert_{\mathcal{R}_{+}^{f,out}}^{2}}=\\
=e^{\Vert\hat{V}_{+}\mathbf{u}_{\eta_{j},r_{j}}^{-}\Vert_{\mathcal{R}_{+}^{f,out}}^{2}\left(1-e^{-2\eta_{j}\tau}(1-s)\right)}=e^{\Vert\mathbf{u}_{\eta_{j},r_{j}}^{-}\Vert_{\mathcal{H}_{p_{0}}^{s}}^{2}\left(1-e^{-2\eta_{j}\tau}(1-s)\right)}
\end{multline*}
The right hand side of the last equation is a generating function
for a pure death process in which the number of quanta in $\Gamma_{s}(\mathcal{H}_{p_{0}}^{s,out})$,
counted by $\lambda(\tilde{P}_{+}^{\perp})$, decays monotonically
over time. Since $\lambda(\tilde{P}_{+})+\lambda(\tilde{P}_{+}^{\perp})=\lambda(I_{\mathcal{R}_{+}^{f,out}})$
and since $\lambda(I_{\mathcal{R}_{+}^{f,out}})$ counts the total
number of quanta in $\Gamma_{s}(\mathcal{R}_{+}^{f,out})$ which is
conserved under the evolution by $\left\{ \Gamma_{0}(\hat{U}(\tau))\right\} _{\tau\geq0}$,
we conclude that the number of quanta counted by $\lambda(\tilde{P}_{+})$,
i.e., the number of quanta in $\Gamma_{s}(\mathcal{D}_{+}^{f,out})$
monotonically increases over time, that is, the quanta emitted by
the system is absorbed in the Fock space $\Gamma_{s}(\mathcal{D}_{+}^{f,out})$.
This shows that the stable motion of $\left\{ \tilde{Z}_{f}(\tau)\right\} _{\tau\geq0}$
is induced by the emission of quanta into the dynamical environment
described by the Fock space $\Gamma_{s}(\mathcal{D}_{+}^{f,out})$.
Similar results are obtained with respect to the isometric dilation
of the unstable semigroup $\left\{ \tilde{Z}_{b}(\tau)\right\} _{\tau\leq0}$
in the space $\mathcal{R}_{+}^{b,out}=\mathcal{H}_{p_{0}}^{u,out}\oplus\mathcal{D}_{+}^{b,out}$
and the corresponding Fock space obtained in the process of second
quantiaztion
\[
\Gamma_{s}(\mathcal{R}_{+}^{b,out})=\Gamma_{s}(\mathcal{H}_{p_{0}}^{u,out})\otimes\Gamma_{s}(\mathcal{D}_{+}^{b,out})
\]
In this case the quanta emitted from $\Gamma_{s}(\mathcal{H}_{p_{0}}^{u,out})$
when applying the (lifting of the) backward isometric evolution is
absorbed in the environment $\Gamma_{s}(\mathcal{D}_{+}^{b,out})$.
If we consider this later emission process in reversed direction of
time, i.e., for forward propagation, we obtain a process of absorption
of quanta from the environment $\Gamma_{s}(\mathcal{D}_{+}^{b,out})$
into the system $\Gamma_{s}(\mathcal{H}_{p_{0}}^{u,out})$, inducing
the instability of the motion associated with the unstable semigroup
$\left\{ \tilde{Z}_{b}(\tau)\right\} _{\tau\leq0}$.

\section{Conclusions\label{sec:conclusions}}

We have studied the stability of the trajectories generated by a Hamiltonian
of the form of Eq. (\ref{eq:geometric_hamiltonian-1}). The local
stability of such a system can be described by the geodesic deviation
associated with these trajectories. It is well known that the analysis
of such a system is based on the identification of the second covariant
derivative of the geodesic deviation vector with the structure of
a harmonic oscillator. This argument is generally based on an implicit
assumption that in the limit of locally flat coordinate system in
the neighborhood of a point the geodesic deviation equation is well
approximated by an oscillator equation. We have shown that this result
can, in fact, be rigorously derived by means of an explicit unitary
map of the geodesic deviation equation into a parametric oscillator
equation. By transcribing this second order parametric oscillator
equation into the corresponding form of a first order dynamical system,
one finds that stable and unstable behavior are clearly separated.
This construction, furthermore, supplies a symplectic form for the
system of dynamical variables, which then lends itself to a second
quantization which permits the identification of the excitation modes
with the dynamical behavior of the system.

Assuming that the curvature tensor entering into the geodesic deviation
equation is slowly changing we locally approximate the behavior of
the solution of the dynamical system corresponding to the geodesic
deviation equation in terms of a forward contractive semigroup for
the stable part of the evolution and a backward contractive semigroup
for the unstable part of the evolution. We then apply a Sz.-Nagy-Foias
dilation procedure to obtain an isometric dilation of both the forward
and backward semigroups. The dilation of these semigroups leads to
an understanding of the dynamical behavior of the system in terms
of an interaction of the system with a field representing an evironment. 

The dilation procedure introduces degrees of freedom associated with
the stability of the system. Second quantization of the dilated system
provides an interpretation of the dynamical behavior of the original
system. The stability of the stable part of the original system is
associated with the emission of quanta into an environment corresponding
to the additional degrees of freedom introduced in the dilation process.
Similarly, the instability of the unstable part of the evolution is
associated with the absorption of quanta.

The structure we have described constitutes an embedding of a conservative
physical system into a larger system with quantized degrees of freedom
which provides an interpretation of the instabilities of the original
system. As in the case of the damped harmonic oscillator \cite{key-7}
where the qunatized degrees of freedom associated with the dilation
may be put into correspondence with radiation due to the friction
in the oscillator, we could imagine that the quantized degrees of
freedom of the instabilities of a dynamical system have observable
consequences which might be seen in intrinsic thermodynamic properties
of the system. The treatment carried out in this work could, moreover,
provide a rigorous framework for the considerations of Kandrup et.
al. \cite{key-8} based on the work of Cassetti et. al. \cite{key-14}
on the association of the behavior of chaotic systems with thermodynamic
properties.

\section*{Appendix A: Unitary and isometric dilations of contractive semigroups}

Let $\left\{ Z(\tau)\right\} _{\tau\in[0,\infty)}$ be a continuous,
strongly contractive, semigroup on a Hilbert space $\mathcal{H}$
satisfying $\text{s-}\lim_{\tau\to\infty}Z(\tau)=0$, i.e., for every
$\psi\in\mathcal{H}$ we have $\lim_{\tau\to\infty}\Vert Z(\tau)\psi\Vert=0$.
According to the Sz.-Nagy-Foias theory of contraction operators and
contractive semigroups on Hilbert space \cite{key-5} for such a semigroup
there exists an\emph{ isometric dilation}, i.e., there exists a Hilbert
space $\mathcal{R}_{+}$, an isometric semigroup $\left\{ U_{+}(\tau)\right\} _{\tau\geq0}$
defined on $\mathcal{R}_{+},$ a subspace $\mathcal{H}_{+}\subset\mathcal{R}_{+}$
and an isometric isomorphism $V_{+}\,:\,\mathcal{H}\mapsto\mathcal{H}_{+}$
such that
\[
Z(\tau)=V_{+}^{*}Z_{+}(\tau)V_{+},\quad\tau\geq0,
\]
where

\[
Z_{+}(\tau):=P_{\mathcal{H}_{+}}U_{+}(\tau)P_{\mathcal{H}_{+}},\quad\tau\geq0,
\]
and $P_{\mathcal{H}_{+}}$ is the orthogonal projection in $\mathcal{R}_{+}$
on the subspace $\mathcal{H}_{+}$. Therefore, for $\tau\geq0$, $Z(\tau)$
is unitarily equivalent to the projection of $U_{+}(\tau)$ onto the
subspace $\mathcal{H}_{+}\subset\mathcal{R}_{+}$ representing $\mathcal{H}$.
The isometric dilation is called \emph{minimal} if $\mathcal{R}_{+}=\overline{\vee_{\tau\geq0}U_{+}(\tau)\mathcal{H}_{+}}$.
We refer to $\mathcal{R}_{+}$ as the dilation Hilbert space for the
isometric dilation and to $\left\{ U_{+}(\tau)\right\} _{\tau\in[0,\infty)}$
as an isometric dilation of the semigroup $\left\{ Z(\tau)\right\} _{\tau\geq0}$.
The dilation Hilbert space $\mathcal{R}_{+}$ is naturally decomposed
into two orthogonal subspaces
\[
\mathcal{R}_{+}=\mathcal{H}_{+}\oplus\mathcal{D}_{+}
\]
and, moreover, $\mathcal{D}_{+}$ is invariant under $U_{+}(\tau)$
for $\tau\geq0$. 

According to the Sz.-Nagy-Foias theory an isometric dilation of the
semigroup $\left\{ Z(\tau)\right\} _{\tau\geq0}$ can be extended
into a unitary dilation of the same semigroup, i.e., there exists
a Hilbert space $\mathcal{R}$ and a unitary evolution group $\left\{ U(\tau)\right\} _{\tau\in\mathbb{R}}$
defined on $\mathcal{R}$, such that $\mathcal{R}_{+}\subset\mathcal{R}$
and for each $\tau\geq0$, $U_{+}(\tau)=\left.U(\tau)\right|_{\mathcal{R}_{+}}$.
Denoting $\mathcal{D}_{-}=\mathcal{R}\ominus\mathcal{R}_{+}$ we have
\begin{equation}
\mathcal{R}=\mathcal{D}_{-}\oplus\mathcal{R}_{+}=\mathcal{D}_{-}\oplus\mathcal{H}_{+}\oplus\mathcal{D}_{+}\label{eq:unitary_dilation_space_decomp}
\end{equation}
and 
\[
Z_{+}(\tau)=P_{\mathcal{H}_{+}}U(\tau)P_{\mathcal{H}_{+}},\quad\tau\geq0
\]
 We refer to $\mathcal{R}$ as the dilation Hilbert space for the
unitary dilation and to $\left\{ U(\tau)\right\} _{\tau\in\mathbb{R}}$
as the unitary dilation of the semigroup $\left\{ Z(\tau)\right\} _{\tau\geq0}$.
The unitary dilation is called \emph{minimal} if $\mathcal{R}=\overline{\vee_{t\in\mathbb{R}}U(\tau)\mathcal{H}_{+}}$.
The minimal isometric and unitary dilations of a given semigroup are
unique up to unitary equivalence. Note that, since $\mathcal{R}_{+}\subset\mathcal{R}$
is invariant under $U(\tau)$ for $\tau\geq0$ then $\mathcal{D}_{-}$
is invariant under $U^{*}(\tau)=U(-\tau)$ for $\tau\geq0$. Thus,
in the decomposition of $\mathcal{R}$ in Eq. (\ref{eq:unitary_dilation_space_decomp})
the subspaces $\mathcal{D}_{-}$ and $\mathcal{D}_{+}$ are stable
under the evolution $U(\tau)$ for $\tau\leq0$ and $\tau\geq0$ respectively.

We construct below two different representations of the unitary and
isometric dilations of a semigroup $\left\{ Z(\tau)\right\} _{\tau\in[0,\infty)}$
with the properties stated above. Denote the generator of $\left\{ Z(\tau)\right\} _{\tau\in[0,\infty)}$
by $B$ and decompose $B$ into $B=B_{+}+iB_{-}$, where $B_{+}:=\frac{1}{2}(B+B^{*})$
and $B_{-}:=\frac{1}{2i}(B-B^{*})$ are self-adjoint so that $Z(\tau)=e^{-iB\tau}=e^{-iB_{+}\tau+B_{-}\tau}$,
$\tau\geq0$. Since the semigroup is strongly contractive and its
strong limit is zero the dissipative part of the generator $B$, i.e.,
the operator $(-B_{-})=\frac{i}{2}(B-B^{*})$ is positive definite,
i.e., we have
\[
(\psi,\,(-B_{-})\psi)_{\mathcal{H}}>0,\quad\forall\psi\in\mathcal{H},\ \psi\not=0
\]
\begin{multline*}
(\psi,\,(-B_{-})\psi)_{\mathcal{H}}=-(\psi,\,\frac{1}{2i}(B-B^{*})\psi)_{\mathcal{H}}=\frac{1}{2i}(\psi,\,(B^{*}-B)\psi)_{\mathcal{H}}=\\
=\frac{1}{2i}(\psi,\, B^{*}\psi)_{\mathcal{H}}-\frac{1}{2i}(\psi,\, B\psi)_{\mathcal{H}}=\frac{1}{2i}(B\psi,\,\psi)_{\mathcal{H}}-\frac{1}{2i}(\psi,\, B\psi)_{\mathcal{H}}=\\
=\frac{1}{2}\left[-i(B\psi,\,\psi)_{\mathcal{H}}+i(\psi,\, B\psi)_{\mathcal{H}}\right]=\frac{1}{2}\left[(iB\psi,\,\psi)_{\mathcal{H}}+(\psi,\, iB\psi)_{\mathcal{H}}\right]=-\frac{1}{2}\left.\left(\frac{d}{d\tau}\Vert Z(\tau)\psi\Vert^{2}\right)\right|_{t=0}>0
\end{multline*}
Let $\tilde{\mathcal{R}}$ be the Hilbert space of all vector valued
functions defined on $\mathbb{R}$ with values in $\mathcal{H}$ and
inner product defined by
\[
\langle f,\, g\rangle_{\tilde{\mathcal{R}}}:=\intop_{-\infty}^{\infty}(f(t),\, i(B-B^{*})g(t))_{\mathcal{H}}\, dt,\qquad f,g\in\tilde{\mathcal{R}}
\]
and with the corresponding norm
\begin{multline*}
\Vert f\Vert_{\tilde{\mathcal{R}}}^{2}=\intop_{-\infty}^{\infty}(f(t),\, i(B-B^{*})f(t))_{\mathcal{H}}\, dt=\intop_{-\infty}^{\infty}(f(t),\,(-2B_{-})f(t))_{\mathcal{H}}\, dt=\\
=\intop_{-\infty}^{\infty}((-2B_{-})^{1/2}f(t),\,(-2B_{-})^{1/2}f(t))_{\mathcal{H}}\, dt=\intop_{-\infty}^{\infty}\Vert(-2B_{-})^{1/2}f(t)\Vert_{\mathcal{H}}^{2}\, dt,\qquad f\in\tilde{\mathcal{R}}
\end{multline*}
On $\tilde{\mathcal{R}}$ define the evolution group $\left\{ \tilde{U}(\tau)\right\} _{\tau\in\mathbb{R}}$
by 
\begin{equation}
[\tilde{U}(\tau)g](t)=g(t-\tau),\qquad g\in\tilde{\mathcal{R}}\,.\label{eq:R_tilde_evolution_group}
\end{equation}
Now define a mapping $\tilde{V}_{+}:\,\mathcal{H}\mapsto\tilde{\mathcal{R}}$
by
\[
[\tilde{V}_{+}\psi](t)=\Theta(-t)Z(-t)\psi=\bigg\{\begin{matrix}Z(-t)\psi, & t\leq0\\
0, & t>0
\end{matrix}\bigg\},\quad t\in\mathbb{R},\ \psi\in\mathcal{H}\,.
\]
We have
\begin{multline*}
\langle\tilde{V}_{+}\phi,\tilde{V}_{+}\psi\rangle_{\tilde{\mathcal{R}}}=i\int_{-\infty}^{\infty}dt\,\,(\Theta(-t)Z(-t)\phi,(B-B^{*})\Theta(-t)Z(-t)\psi)_{\mathcal{H}}=\\
=i\int_{-\infty}^{0}dt\,\,(Z(-t)\phi,(B-B^{*})Z(-t)\psi)_{\mathcal{H}}=i\intop_{0}^{\infty}dt\,(Z(t)\phi,(B-B^{*})Z(t)\psi)_{\mathcal{H}}=\\
=\intop_{0}^{\infty}dt\,[(Z(t)\phi,iBZ(t)\psi)_{\mathcal{H}}+(iBZ(t)\phi,Z(t)\psi)_{\mathcal{H}}]=-\intop_{0}^{\infty}dt\frac{d}{dt}(Z(t)\phi,Z(t)\psi)=\\
=(Z(0)\phi,Z(0)\psi)_{\mathcal{H}}=(\phi,\psi)_{\mathcal{H}}
\end{multline*}
Clearly $\tilde{V}_{+}$ is one to one and, hence, $\tilde{V}_{+}$
defines a unitary embedding of $\mathcal{H}$ into $\tilde{\mathcal{R}}$.
Moreover, $\left\{ \tilde{U}(\tau)\right\} _{\tau\in\mathbb{R}}$
is a unitary dilation of the semigroup $\left\{ Z(\tau)\right\} _{\tau\in[0,\infty)}$
with respect to this embedding and $\tilde{\mathcal{R}}$ is the dilation
Hilbert space. Indeed, for any $\tau\geq0$ we have
\begin{multline*}
\langle\tilde{V}_{+}\phi,\tilde{U}(\tau)\tilde{V}_{+}\psi\rangle_{\tilde{\mathcal{R}}}=\intop_{-\infty}^{\infty}dt\,([\tilde{V}_{+}\phi](t),i(B-B^{*})[\tilde{U}(\tau)(\tilde{V}_{+}\psi)](t))_{\mathcal{H}}=\\
=\intop_{-\infty}^{\infty}dt\,([\tilde{V}_{+}\phi](t),i(B-B^{*})[\tilde{V}_{+}\psi](t-\tau))_{\mathcal{H}}=\\
=\intop_{-\infty}^{\infty}dt\,(\Theta(-t)Z(-t)\phi,i(B-B^{*})\Theta(-t+\tau)Z(-t+\tau)\psi)_{\mathcal{H}}=\\
=\intop_{-\infty}^{0}dt\,(Z(-t)\phi,i(B-B^{*})Z(-t+\tau)\psi)_{\mathcal{H}}=\intop_{0}^{\infty}dt\,(Z(t)\phi,i(B-B^{*})Z(t+\tau)\psi)_{\mathcal{H}}=\\
=\intop_{0}^{\infty}dt\,(Z(t)\phi,i(B-B^{*})Z(t)Z(\tau)\psi)_{\mathcal{H}}=(\phi,Z(\tau)\psi)_{\mathcal{H}}
\end{multline*}
Now define a mapping $L:\,\tilde{\mathcal{R}}\mapsto L^{2}(\mathbb{R};\mathcal{H})$
\begin{equation}
(Lg)(t)\,:=(-2B_{-})^{1/2}g(t),\quad g\in\tilde{\mathcal{R}},\ t\in\mathbb{R}\,.\label{eq:L_map}
\end{equation}
With this definition we have
\begin{multline*}
\langle g,g'\rangle_{\tilde{\mathcal{R}}}:=\int_{-\infty}^{\infty}dt\,\,(g(t),\, i(B-B^{*})g'(t))_{\mathcal{H}}=\int_{-\infty}^{\infty}dt\,\,(g(t),(-2B_{-})g'(t))_{\mathcal{H}}=\\
=\int_{-\infty}^{\infty}dt\,\,((-2B_{-})^{1/2}g(t),(-2B_{-})^{1/2}g'(t))_{\mathcal{H}}=\int_{-\infty}^{\infty}dt\,\,([Lg](t),[Lg'](t))_{\mathcal{H}}=\langle Lg,Lg'\rangle_{L^{2}(\mathbb{R},\mathcal{H})}
\end{multline*}

\noindent Let $\hat{U}(\tau)$ be the transformation of the evolution
$\tilde{U}(\tau)$ by $L$ from $\tilde{\mathcal{R}}$ into $L^{2}(\mathbb{R};\mathcal{H})$.
We have 
\[
[\hat{U}(\tau)Lg](t)=[L\tilde{U}(\tau)g](t)=(-2B_{-})^{1/2}[\tilde{U}(\tau)g](t)=(-2B_{-})^{1/2}g(t-\tau)=[Lg](t-\tau)
\]
so that $\hat{U}(\tau)$ is again translation. We continue to construct
an embedding of $\mathcal{H}$ into $L^{2}(\mathbb{R};\mathcal{H})$
via a map $\hat{V}_{+}:\,\mathcal{H}\mapsto L^{2}(\mathbb{R};\mathcal{H})$
defined by 
\[
\hat{V}_{+}:=L\tilde{V}_{+}.
\]
Indeed we have
\[
\langle\hat{V}_{+}\phi,\hat{V}_{+}\psi\rangle_{L^{2}(\mathbb{R};\mathcal{H})}=\langle L\tilde{V}_{+}\phi,L\tilde{V}_{+}\psi\rangle_{L^{2}(\mathbb{R};\mathcal{H})}=\langle\tilde{V}_{+}\phi,\tilde{V}_{+}\psi\rangle_{\tilde{\mathcal{R}}}=(\phi,\psi)_{\mathcal{H}}
\]
The mapping $\hat{V}_{+}$ is given explicitely by
\[
[\hat{V}_{+}\psi](t)=[LV_{+}\psi](t)=(-2B_{-})^{1/2}\Theta(-t)Z(-t)\psi=\begin{cases}
(-2B_{-})^{1/2}Z(-t)\psi, & t\leq0\\
0 & t>0
\end{cases}
\]
One can easily verify that with the use of the mapping $\hat{V}_{+}$
and the evolution $\left\{ \hat{U}(\tau)\right\} _{\tau\in\mathbb{R}}$
we again obtain a dilation of the semigroup $\left\{ Z(\tau)\right\} _{\tau\in[0,\infty)}$
, i.e.,
\[
\langle\hat{V}_{+}\phi,\,\hat{U}(\tau)\hat{V}_{+}\psi\rangle_{L^{2}(\mathbb{R};\mathcal{H})}=(\phi,\, Z(\tau)\psi)_{\mathcal{H}},\quad\tau\geq0\,,
\]
thus, $\left\{ \hat{U}(\tau)\right\} _{\tau\in\mathbb{R}}$ is a unitary
dilation of $\left\{ Z(\tau)\right\} _{\tau\in[0,\infty)}$ in the
dilation Hilbert space $L^{2}(\mathbb{R};\mathcal{H})$. We call this
representation of the unitary dilation of $\left\{ Z(\tau)\right\} _{\tau\in[0,\infty)}$
in $L^{2}(\mathbb{R};\mathcal{H})$ the \emph{outgoing representation}.
If a unitary dilation of the semigroup $\left\{ Z(\tau)\right\} _{\tau\in[0,\infty)}$
is given by a unitary evolution group $\left\{ U(\tau)\right\} _{\tau\in\mathbb{R}}$
defined on a dilation Hilbert space $\mathcal{R}$ with an embedding
of $\mathcal{H}$ into $\mathcal{R}$ given by a mapping $V_{+}$
then we denote by $\hat{W}_{+}\,:\,\mathcal{R}\mapsto L^{2}(\mathbb{R},\mathcal{H})$
the unitary mapping of the unitary dilation in $\mathcal{R}$ onto
the outgoing representation. Denoting
\[
\mathcal{D}_{-}^{out}:=\hat{W}_{+}\mathcal{D}_{-},\qquad\mathcal{H}^{out}:=\hat{V}_{+}\mathcal{H}=\hat{W}_{+}\mathcal{H}_{+},\qquad\mathcal{D}_{+}^{out}:=\hat{W}_{+}\mathcal{D}_{+},
\]
we have
\[
\mathcal{D}_{-}^{out}=L^{2}(\mathbb{R}_{-},\mathcal{H})\ominus\mathcal{H}^{out},\qquad\mathcal{H}^{out}\subset L^{2}(\mathbb{R}_{-},\mathcal{H}),\qquad\mathcal{D}_{+}^{out}=L^{2}(\mathbb{R}_{+},\mathcal{H}),
\]
so that
\[
L^{2}(\mathbb{R},\mathcal{H})=\mathcal{D}_{-}^{out}\oplus\mathcal{H}^{out}\oplus\mathcal{D}_{+}^{out}
\]
Moreover, we have
\[
\hat{U}(\tau)=\hat{W}_{+}U(\tau)\hat{W}_{+}^{-1},
\]
so that $U(\tau)$ is represented in this represntation by translation
to the right by $\tau$ units. We observe that the isometric dilation
of $\left\{ Z(\tau)\right\} _{\tau\in[0,\infty)}$ is obtained by
projection from $\mathcal{R}$ onto $\mathcal{R}_{+}$. If the projection
onto the subspace $\mathcal{D}_{-}\subset\mathcal{R}$ is denoted
by $P_{-}$ and we denote $P_{-}^{\perp}=I-P_{-}$, then the Hilbert
space for the isometric dilation is given by $\mathcal{R}_{+}=P_{-}^{\perp}\mathcal{R}$
and the isometric semigroup $\left\{ U_{+}(\tau)\right\} _{\tau\in[0,\infty)}$,
dilating $\left\{ Z(\tau)\right\} _{\tau\in[0,\infty)}$ in $\mathcal{R}_{+}$,
is given by $U_{+}(\tau)=P_{-}^{\perp}U(\tau)P_{-}^{\perp}$. In the
outgoing representation $\mathcal{R}_{+}$ is represented by a subspace
$\mathcal{R}_{+}^{out}\subset L^{2}(\mathbb{R},\mathcal{H})$ given
by
\[
\mathcal{R}_{+}^{out}=\mathcal{H}^{out}\oplus\mathcal{D}_{+}^{out}
\]
and $\left\{ U_{+}(\tau)\right\} _{\tau\in[0,\infty)}$ is represented
by
\[
U_{+}^{out}(\tau)=P_{-}^{\perp,out}U(\tau)P_{-}^{\perp,out}=U(\tau)P_{-}^{\perp,out},\quad\tau\geq0
\]
where the second equality is due to the invariance of $\mathcal{R}_{+}^{out}$
under $U(\tau)$ for $\tau\geq0$.

We now construct a second representation, called the \emph{incoming
representation,} of the unitary dilation of $\left\{ Z(\tau)\right\} _{\tau\in[0,\infty)}$.
The dilation Hilbert space in this representation is again $L^{2}(\mathbb{R},\mathcal{H})$.
Consider first the Hilbert space $\tilde{\mathcal{R}}$ defined above
and the evolution group $\left\{ \tilde{U}(\tau)\right\} _{\tau\in\mathbb{R}}$
defined on $\tilde{\mathcal{R}}$ in Eq. (\ref{eq:R_tilde_evolution_group}).
Define a mapping $V_{-}:\,\mathcal{H}\mapsto\tilde{\mathcal{R}}$
by
\[
[V_{-}\psi](t)=\Theta(t)Z^{*}(t)\psi=\bigg\{\begin{matrix}0, & t<0\\
Z^{*}(t)\psi, & t\geq0
\end{matrix}\bigg\},\quad t\in\mathbb{R},\ \psi\in\mathcal{H}\,.
\]
We have
\begin{multline*}
\langle V_{-}\phi,V_{-}\psi\rangle_{\tilde{\mathcal{R}}}=\int_{-\infty}^{\infty}dt\,\,(\Theta(t)Z^{*}(t)\phi,i(B-B^{*})\Theta(t)Z^{*}(t)\psi)_{\mathcal{H}}=\\
=\int_{0}^{\infty}dt\,\,(Z^{*}(t)\phi,i(B-B^{*})Z^{*}(t)\psi)_{\mathcal{H}}=\\
=-\intop_{0}^{\infty}dt\,[(Z^{*}(t)\phi,iB^{*}Z^{*}(t)\psi)_{\mathcal{H}}+(iB^{*}Z(t)\phi,Z(t)\psi)_{\mathcal{H}}]=-\intop_{0}^{\infty}dt\frac{d}{dt}(Z^{*}(t)\phi,Z^{*}(t)\psi)=\\
=(Z^{*}(0)\phi,Z^{*}(0)\psi)_{\mathcal{H}}=(\phi,\psi)_{\mathcal{H}}
\end{multline*}
Clearly $V_{-}$ is one to one and, hence, $V_{-}$ is a unitary embedding
of $\mathcal{H}$ into $\tilde{\mathcal{R}}$. Moreover, $\left\{ \tilde{U}(\tau)\right\} _{\tau\in\mathbb{R}}$
is a unitary dilation of the semigroup $\left\{ Z(\tau)\right\} _{\tau\in[0,\infty)}$
with respect to this embedding and $\tilde{\mathcal{R}}$ is the dilation
Hilbert space. Indeed, for any $\tau\geq0$ we have
\begin{multline*}
\langle V_{-}\phi,\tilde{U}(\tau)V_{-}\psi\rangle_{\tilde{\mathcal{R}}}=\intop_{-\infty}^{\infty}dt\,([V_{-}\phi](t),i(B-B^{*})[\tilde{U}(\tau)V_{-}\psi](t))_{\mathcal{H}}=\\
=\intop_{-\infty}^{\infty}dt\,([V_{-}\phi](t),i(B-B^{*})[V_{-}\psi](t-\tau))_{\mathcal{H}}=\\
=\intop_{-\infty}^{\infty}dt\,(\Theta(t)Z^{*}(t)\phi,i(B-B^{*})\Theta(t-\tau)Z^{*}(t-\tau)\psi)_{\mathcal{H}}=\\
=\intop_{\tau}^{\infty}dt\,(Z^{*}(t)\phi,i(B-B^{*})Z^{*}(t-\tau)\psi)_{\mathcal{H}}=\intop_{0}^{\infty}d\tilde{t}\,(Z^{*}(\tilde{t}+\tau)\phi,i(B-B^{*})Z^{*}(\tilde{t})\psi)_{\mathcal{H}}=\\
=\intop_{0}^{\infty}d\tilde{t}\,(Z^{*}(\tilde{t})Z^{*}(\tau)\phi,i(B-B^{*})Z^{*}(\tilde{t})\psi)_{\mathcal{H}}=(Z^{*}(\tau)\phi,\psi)_{\mathcal{H}}=(\phi,Z(\tau)\psi)_{\mathcal{H}}
\end{multline*}
We continue to construct an embedding of $\mathcal{H}$ into $L^{2}(\mathbb{R};\mathcal{H})$
via a map $\hat{V}_{-}:\,\mathcal{H}\mapsto L^{2}(\mathbb{R};\mathcal{H})$
defined by 
\[
\hat{V}_{-}:=LV_{-}
\]
where $L:\,\tilde{\mathcal{R}}\mapsto L^{2}(\mathbb{R};\mathcal{H})$
is the mapping defined above in Eq. (\ref{eq:L_map}). Indeed we have
\[
\langle\hat{V}_{-}\phi,\hat{V}_{-}\psi\rangle_{L^{2}(\mathbb{R};\mathcal{H})}=\langle LV_{-}\phi,LV_{-}\psi\rangle_{L^{2}(\mathbb{R};\mathcal{H})}=\langle V_{-}\phi,V_{-}\psi\rangle_{\tilde{\mathcal{R}}}=(\phi,\psi)_{\mathcal{H}}
\]
The mapping $\hat{V}_{-}$ is given explicitely by
\[
[\hat{V}_{-}\psi](t)=[LV_{-}\psi](t)=(-2B_{-})^{1/2}\Theta(t)Z^{*}(t)\psi=\begin{cases}
0, & t<0\\
(-2B_{-})^{1/2}Z^{*}(t)\psi, & t\geq0
\end{cases}
\]

\noindent Leting $\hat{U}(\tau)$ be the transformation of the evolution
$\tilde{U}(\tau)$ by $L$ from $\tilde{\mathcal{R}}$ into $L^{2}(\mathbb{R};\mathcal{H})$,
as above, one can verify that with the use of the mapping $\hat{V}_{-}$
and the evolution $\left\{ \hat{U}(\tau)\right\} _{a\in\mathbb{R}}$
we again obtain a unitary dilation of the semigroup $\left\{ Z(\tau)\right\} _{\tau\in[0,\infty)}$
, i.e.,
\[
\langle\hat{V}_{-}\phi,\,\hat{U}(\tau)\hat{V}_{-}\psi\rangle_{L^{2}(\mathbb{R};\mathcal{H})}=(\phi,\, Z(\tau)\psi)_{\mathcal{H}},\quad\tau\geq0.
\]
Let a unitary dilation of the semigroup $\left\{ Z(\tau)\right\} _{\tau\in[0,\infty)}$
be given by a unitary evolution group $\left\{ U(\tau)\right\} _{\tau\in\mathbb{R}}$
defined on a dilation Hilbert space $\mathcal{R}$ with an embedding
of $\mathcal{H}$ into $\mathcal{R}$ given by a mapping $V_{+}$.
We denote by $\hat{W}_{-}\,:\,\mathcal{R}\mapsto L^{2}(\mathbb{R},\mathcal{H})$
the unitary mapping of the unitary dilation in $\mathcal{R}$ onto
the incoming representation. Denoting
\[
\mathcal{D}_{-}^{in}:=\hat{W}_{-}\mathcal{D}_{-},,\qquad\mathcal{H}^{in}:=\hat{V}_{-}\mathcal{H}=\hat{W}_{-}\mathcal{H}_{+},\qquad\mathcal{D}_{+}^{in}:=\hat{W}_{-}\mathcal{D}_{+}
\]
we have
\[
\mathcal{D}_{-}^{in}=L^{2}(\mathbb{R}_{-},\mathcal{H}),\qquad\mathcal{H}^{in}\subset L^{2}(\mathbb{R}_{+},\mathcal{H}),\qquad\mathcal{D}_{+}^{in}=L^{2}(\mathbb{R}_{+},\mathcal{H})\ominus\mathcal{H}^{in},
\]
so that
\[
L^{2}(\mathbb{R},\mathcal{H})=\mathcal{D}_{-}^{in}\oplus\mathcal{H}^{in}\oplus\mathcal{D}_{+}^{in}
\]
Moreover, we have
\[
\hat{U}(\tau)=\hat{W}_{-}U(\tau)\hat{W}_{-}^{-1},
\]
so that $U(\tau)$ is represented by translation to the right by $\tau$
units. We observe that in the incoming representation $\mathcal{R}_{+}$
is represented by a subspace $\mathcal{R}_{+}^{in}\subset L^{2}(\mathbb{R},\mathcal{H})$
given by
\[
\mathcal{R}_{+}^{in}=\mathcal{H}^{in}\oplus\mathcal{D}_{+}^{in}=L^{2}(\mathbb{R}_{+},\mathcal{H})
\]
and $\left\{ U_{+}(\tau)\right\} _{\tau\in[0,\infty)}$ is represented
by
\[
U_{+}^{in}(\tau)=P_{-}^{\perp,in}U(\tau)P_{-}^{\perp,in}=U(\tau)P_{-}^{\perp,in},\quad\tau\geq0
\]
where the second equality is due to the invariance of $\mathcal{R}_{+}^{in}$
under $U(\tau)$ for $\tau\geq0$.

\end{document}